\documentclass[apj]{emulateapj}

\usepackage{amsmath}
\usepackage{graphics}
\usepackage{epsfig}
\usepackage{graphicx}
\usepackage{color}
\usepackage{epstopdf}
\usepackage{epsfig}
\usepackage{natbib}
\usepackage[percent]{overpic}

\newcommand{\etal}{et~al.\ }
\newcommand{\Ha}{H$\alpha$ }
\newcommand{\Hans}{H$\alpha$} 

\newcommand{\msun}{M_{\odot}}

\def\msun{$M_\odot$}

\shorttitle{SED fits of BCDs}
\shortauthors{Janowiecki \etal}

\received{ 15 Sep 2016 }
\accepted{ 12 Dec 2016 }

\begin{document}

\title{Constraining the Stellar Populations and Star
  Formation Histories of Blue Compact Dwarf Galaxies with SED
  fits}

\author{Steven Janowiecki\altaffilmark{1,2},
John J. Salzer\altaffilmark{2},
Liese van~Zee\altaffilmark{2},
Jessica L. Rosenberg\altaffilmark{3},
Evan Skillman\altaffilmark{4}
}
\email{steven.janowiecki@uwa.edu.au
}

\altaffiltext{1}{International Center for Radio Astronomy Research,
  M468, The University of Western Australia, 35 Stirling Highway,
  Crawley, Western Australia, 6009, Australia}
\altaffiltext{2}{Department of Astronomy, Indiana University, 727 East
  Third Street, Bloomington, IN 47405, USA}
\altaffiltext{3}{Department of Physics and Astronomy, George Mason University, Fairfax, VA 22030, USA}
\altaffiltext{4}{Minnesota Institute for Astrophysics, University of Minnesota, 116 Church Street, SE Minneapolis, MN, 55455 USA}


\begin{abstract}

We discuss and test possible evolutionary connections
between Blue Compact Dwarf galaxies (BCDs) and other types of dwarf
galaxies. BCDs provide ideal laboratories to study intense star
formation episodes in low mass dwarf galaxies, and have sometimes been
considered a short-lived evolutionary stage between types of dwarf
galaxies. To test these connections, we consider a sample of BCDs as
well as a comparison sample of nearby galaxies from the Local Volume
Legacy (LVL) survey for context. We fit the multi-wavelength spectral
energy distributions (SED, far-ultra-violet to far-infrared) of each
galaxy with a grid of theoretical models to determine their stellar
masses and star formation properties. 
We compare our results for BCDs with the LVL galaxies to
put BCDs in the context of normal galaxy evolution. The SED fits
demonstrate that the star formation events currently underway in BCDs
are at the extreme of the continuum of normal dwarf galaxies, both in
terms of the relative mass involved and in 
the relative increase over previous star formation
rates. Today's BCDs are distinctive objects
in a state of extreme star formation which is rapidly transforming
them. This study also suggests ways to identify former BCDs whose star
formation episodes have since faded.

\end{abstract}


\section{Introduction}







The formation and evolutionary pathways of dwarf galaxies are still
not well-understood, despite the fact that they are fairly simple
systems and are the most abundant type of galaxy in the universe. Many
studies have focused on the idea that dwarf galaxy evolution is driven by
interactions and mergers, through ram pressure stripping, tidal
disruption, or other external transformations. Today, there are many
well-established mechanisms to evolve dwarf galaxies in high density
group and cluster environments. It has been more difficult to study
the ongoing secular dwarf galaxy evolution in isolated
environments where mergers and interactions have a smaller
impact. In this regime, star formation is the most transformative
process a dwarf galaxy can undergo.

The best places to study the transformative effects of star formation
are in the dwarf galaxies with the highest star formation rates (SFRs): the
Blue Compact Dwarf galaxies (BCDs). Dwarf galaxies are often broadly
categorized as either dwarf irregulars (dIs) which lack a regular
morphology, possess substantial gas reservoirs, and are often forming
stars, or as dwarf spheroidals (dSphs, or dwarf ellipticals, dEs),
which have regular isophotes, are gas-poor, and are not forming
stars. BCDs are classified in a variety of ways, but are
observationally remarkable for
their strong emission lines and star formation
\citep{sargentsearle70, searlesargent72, izotovthuan04}, 
their underlying old stellar population
\citep{loosethuan86, aloisi07},
their exceptionally low gas-phase metallicity
\citep{terlevich91, izotov94, izotovthuan99, hunter99},
and their compact underlying stellar and HI distributions
\citep{papaderos96, vanzee98, janowiecki14, lelli14}. Some groups have
hypothesized that a compact mass distribution is the most
distinctive characteristic of a BCD \citep{vanzee98, lelli14,
  mcquinn15c}.

Recently, many groups have used observations to study the evolutionary
connections between BCDs and other dwarf galaxies. 
Detailed studies of individual extreme objects \citep[e.g.,][]{guseva01,
  ashley14} and systematic surveys of many objects
\citep[e.g.,][]{noeske07} have both suggested that BCDs are members of
a rapidly evolving class of galaxies, and play an important role in
dwarf galaxy evolution. Deep surface photometry studies
\citep{noeske03, noeske05} in addition to spectroscopic observations
\citep{papaderos08} have highlighted the possible connections between
BCDs and dwarf irregulars in terms of their structural parameters and
chemical evolution. 
Other observations suggest that the unusually low metallicity of BCDs
is more likely maintained by outflows of enriched winds
\citep{carigi95, maclow99, mcquinn15b} rather than pristine gas infall
\citep{matteucci83}. There have also been efforts to find galaxies
which may have experienced a BCD-like star formation event at some
point in their history, or to predict the future evolutionary state of
today's BCDs \citep{sanchezalmeida08, sanchezalmeida09, amorin12,
  lelli12, koleva13, meyer14}.
%
%
%
%
%
%

One of the big questions about BCDs is the reason for their
intense star formation, and whether they have been triggered in some
way. 
There
are well-studied examples where mergers or interactions between dIs
have externally triggered intense 
star formation \citep[e.g., II~Zw~40,][]{sargentsearle70, baldwin82,
  terlevich91, vanzee98, bordalo09}. Recently, cosmologically-relevant
hydrodynamical 
simulations are beginning to 
approach these questions from a theoretical perspective
\citep[e.g.,][and references therein]{valcke08}. While 
dwarf-dwarf galaxy mergers become 
increasingly rare at lower masses \citep{deason14}, some groups have
had success in producing BCD-like galaxies through these interactions
\citep{bekki08}. Others have considered the effects of in-spiraling
star-forming clumps in dwarf 
galaxies \citep{elmegreen12}, or the interaction between a dwarf
galaxy and an infalling cloud of gas \citep{verbeke14}, both of which
can reproduce some of the observed properties of BCDs. 
Both the merger
and gas infall scenarios frequently result in substantial structural
changes to the simulated galaxies, and represent extreme
transformations in the life of a dI.
However, in this work we 
focus on the internal secular evolution of isolated BCDs, and avoid
discussing mergers and interactions.

Compared with typical dIs, BCDs have been found to be especially
compact, in their underlying stellar and HI distributions
\citep{papaderos96, vanzee98, janowiecki14, lelli14}. This compactness
may be related to their ability to host such intense starbursts. When
parametrizing the strength of a star-forming event, the birthrate
parameter ($b$$=$SFR$/$$\langle$SFR$\rangle$, see
Section~\ref{bstrength}) is often used, which 
compares the current 
star formation with the lifetime average. In the local universe,
intense star-forming galaxies are rare as only $\sim$$1\%$ of
star-forming galaxies are considered starbursts with $b\ge3$ 
\citep{bergvall15}. When combining this rarity with the stochastic
nature of star-formation in dwarf galaxies \citep[e.g.,][]{lee09,
  weisz12, bauer13}, it would appear that dwarf
galaxies can experience increases and decreases in their star formation
rate, and only a small fraction are starbursting at any given
time. Perhaps the especially-compact BCDs are able to burst more
effectively than typical dIs, and so can reach higher SFR during their
periods of starburst.

To compare the star formation properties and evolutionary history of
BCDs and dIs from the Local Volume Legacy survey
\citep[LVL,][]{dale09}, we consider the wealth of information  
contained in their stellar populations. An understanding of these
stellar populations can constrain the amount and impact of recent and
past star-formation activity. We have obtained multi-wavelength
photometry in order to fit the spectral energy distributions (SEDs)
for the BCDs and LVL galaxies. SED fitting has recently become a
widely-used tool to derive physical 
properties of galaxies, including their stellar masses and star
formation rates \citep[cf.,][]{walcher11}.

Given that stellar mass is often considered the most fundamental
parameter driving a galaxy's evolutionary path \citep[e.g.,][and
  references therein]{tremonti04, kewleyellison08}, 
star forming galaxies are often plotted on a ``main sequence'',
analogously to the main sequence of stellar evolution. This
correlation between the stellar masses and SFRs of star-forming
galaxies has been observed for massive galaxies in the 
nearby universe \citep{brinchmann04, salim07} and at higher redshift
\citep{daddi07}, with a scatter of only $\sim$$0.2$~dex. The fact that
this relationship exists with such low scatter across a wide range of
redshifts seems to imply a universal mode of 
star formation, from which galaxies rarely deviate
\citep{noeske07}. At smaller masses (below $10^9$\msun) the scatter
becomes larger, as episodic bursts of star formation can affect dwarf
galaxies more significantly \citep[e.g.,][]{cook14b, mcquinn10a,
  weisz11}.

Galaxy stellar masses are most commonly determined by converting
observed luminosities to masses \citep[e.g.,][]{bell01,
  McGaughSchombert14}. Often a color (e.g., B$-$V), is used to
determine the mass-to-light ratio, but these estimates become less
reliable for galaxies which deviate from mean scaling relations
(e.g., if they are currently experiencing a starburst). In order to
more uniformly determine stellar masses across our sample of BCDs and
LVL galaxies, a full multi-wavelength SED-fit is necessary.

Our SED fits can be also used to make crude
estimates of the SFR and star formation histories (SFHs) for BCDs and
LVL galaxies, which describe the 
amount of star formation they experienced throughout their
lifetimes. Different observational indicators are 
sensitive to star formation of different ages,
and the SED fits
incorporate all of the multi-wavelength information into a single
best fitting SFH. While the broadband photometry can never produce
as accurate and well-constrained a SFH as resolved stellar photometry
\citep[e.g.,][]{tolstoy09, mcquinn10a, mcquinn15a}, it can still be
useful in  comparing BCDs with typical LVL galaxies.
In particular, 
the stellar masses and SFRs from the SED fits allow us to quantify how
extreme the BCDs are compared with typical LVL galaxies, and to
constrain the possible evolutionary connections between BCDs and other
dwarf galaxies.

This paper is organized as follows. In Section~\ref{obs}, we describe
the multi-wavelength photometric observations of the BCDs and
comparison 
samples. In Section~\ref{cigale} we discuss our SED-fitting methods and
the consistency checks and verifications of our results. In
Section~\ref{sed_results} we show the results of our best SED 
fits. Section~\ref{sed_discussion} contains a discussion of these results
and their implications, and we briefly summarize our main findings in
Section~\ref{sed_summary}.

\begin{deluxetable*}{cccccccccccc}
\centering
\tabletypesize{\small} 
\tablecaption{Photometric observations of the BCD sample
\label{bcdobs}   
}
\tablewidth{14cm}
\tablehead{
%
%
\colhead{Galaxy} & \colhead{\textit{GALEX}} & 
\colhead{U} & \colhead{B} &  \colhead{V} & \colhead{R} & 
\colhead{I} & \colhead{\Ha} & \colhead{J} &  \colhead{H} &
\colhead{K$_s$} & \colhead{Spitzer} }
\startdata
UM~323   & F/N &     & SN/W  & W    & W     & SN  & SN   & SN/W & SN/W & W    &    \\
UM~408   & F/N &     & SN/W  & W    & W     & SN  & SN   & SN/W & SN/W & W    &    \\
Mk~600   & F/N &     & SN/W  & W    & W     & SN  & SN   & SN/W & SN/W & SN/W & I/M \\
II~Zw~40 & F/N & SN  & SN    & SN   & SN    & SN  & SN   & SN   & SN   & SN   & I/M \\
Mk~5     & F/N & SN  & SN/W  & SN   & W     & SN  & SN   & SN   & SN   & SN   & I/M \\
CG~10    & F/N &     & W     & W    & W     &     &      & W    & W    & W    &    \\
I~Zw~18  & F/N & SN  & SN    & SN   & SN    &     & SN   & SN   & SN   & SN   & I/M \\
Was~5    & F/N & SN  & SN    & SN   & W     &     & SN   & SN   & SN   & SN   &    \\
Mk~36    & F/N & SN  & SN    & SN   & W     &     & SN   & SN   & SN   & SN   & I/M \\
UM~439   & F/N & SN  & SN    & SN   & W     &     & SN   & SN   & SN   & SN   &    \\
Mk~750   &     & SN  & SN    & SN   & W     &     & SN   & SN   & SN   & SN   &    \\
UM~461   &  N  & SN  & SN    & SN   & W     &     & SN   & SN   & SN   & SN   & I/M \\
UM~462   & F/N & SN  & SN    & SN   & W     &     & SN   & SN   & SN   & SN   & I/M \\
Mk~67    & F/N & SN  & SN    & SN   & SN    & SN  & SN   & SN   & SN   & SN   & I  \\
Mk~475   & F/N & SN  & SN    & SN   & W     &     & SN   & SN   & SN   & SN   & I/M \\
Mk~900   & F/N &     & SN/W  & W    & W     & SN  & SN   & SN   & SN   & SN   & I/M \\
Mk~324   & F/N & SN  & SN/W  & SN   & SN/W  & SN  & SN   & SN   & SN   & SN   &    \\
Mk~328   & F/N & SN  & SN/W  & SN   & SN/W  & SN  & SN   & SN   & SN   & SN   &    \\
\enddata
\tablecomments{
F/N - FUV and NUV \textit{GALEX} observations;
SN - \citet{sudarsky95} \& \citet{norton97};
W - WIYN 3.5m observations;
I-IRAC, M-MIPS;
}
\end{deluxetable*}

\begin{turnpage}
\begin{deluxetable*}{lrrrrrrrrrrrrrrrrr}
\tabletypesize{\scriptsize}
\tablecaption{Observed fluxes of the BCD sample
\label{bcddata}   
}
\tablewidth{0cm}
\tablehead{
%
%
\colhead{Galaxy} & \colhead{$f_\textrm{FUV}$} & \colhead{$f_\textrm{NUV}$} &
\colhead{$f_\textrm{U}$} & \colhead{$f_\textrm{B}$} &  \colhead{$f_\textrm{V}$} & 
\colhead{$f_\textrm{R}$} & \colhead{$f_\textrm{I}$} & \colhead{$f_\textrm{J}$} &  
\colhead{$f_\textrm{H}$} & \colhead{$f_\textrm{K$_s$}$} & 
\colhead{$f_{3.6\mu}$} & \colhead{$f_{4.5\mu}$} & \colhead{$f_{5.8\mu}$} & \colhead{$f_{8.0\mu}$} & 
\colhead{$f_{24\mu}$} & \colhead{$f_{70\mu}$} & \colhead{$f_{160\mu}$} \\
\colhead{} & \colhead{$\sigma_\textrm{FUV}$} & \colhead{$\sigma_\textrm{NUV}$} &
\colhead{$\sigma_\textrm{U}$} & \colhead{$\sigma_\textrm{B}$} &  \colhead{$\sigma_\textrm{V}$} & 
\colhead{$\sigma_\textrm{R}$} & \colhead{$\sigma_\textrm{I}$} & \colhead{$\sigma_\textrm{J}$} &  
\colhead{$\sigma_\textrm{H}$} & \colhead{$\sigma_\textrm{K$_s$}$} & 
\colhead{$\sigma_{3.6\mu}$} & \colhead{$\sigma_{4.5\mu}$} & \colhead{$\sigma_{5.8\mu}$} & \colhead{$\sigma_{8.0\mu}$} & 
\colhead{$\sigma_{24\mu}$} & \colhead{$\sigma_{70\mu}$} & \colhead{$\sigma_{160\mu}$} \\
\colhead{}         & \colhead{[mJy]} & \colhead{[mJy]} &
\colhead{[mJy]} & \colhead{[mJy]} & \colhead{[mJy]} & 
\colhead{[mJy]} & \colhead{[mJy]} & \colhead{[mJy]} &  
\colhead{[mJy]} & \colhead{[mJy]} & 
\colhead{[mJy]} & \colhead{[mJy]} & \colhead{[mJy]} & \colhead{[mJy]} & 
\colhead{[mJy]} & \colhead{[mJy]} & \colhead{[mJy]} 
%
}
\startdata
UM~323 &   0.712 &   0.700 & \nodata  &   1.445 &   1.540 &   1.670 &   1.864 &   2.638 &   2.338 &   1.900 & \nodata  & \nodata  & \nodata  & \nodata  & \nodata  & \nodata  & \nodata  \\
      &  \nodata   &   0.001 & \nodata  &   0.013 &   0.040 &   0.062 &   0.017 &   0.165 &   0.125 &   0.258 & \nodata  & \nodata  & \nodata  & \nodata  & \nodata  & \nodata  & \nodata  \\
UM~408 &   0.153 &   0.179 & \nodata  &   0.361 &   0.420 &   0.437 &   0.484 &   0.492 &   0.472 & \nodata  & \nodata  & \nodata  & \nodata  & \nodata  & \nodata  & \nodata  & \nodata  \\
      &   0.003 &   0.002 & \nodata  &   0.003 &   0.037 &   0.039 &   0.008 &   0.040 &   0.031 & \nodata  & \nodata  & \nodata  & \nodata  & \nodata  & \nodata  & \nodata  & \nodata  \\
Mk~600 &   0.792 &   0.924 & \nodata  &   2.524 &   2.788 &   3.051 &   3.296 &   3.978 &   3.985 &   2.647 &   2.083 &   1.879 &   1.680 &   1.440 &  21.568 &   161.2 &   139.6 \\
      &   0.005 &   0.004 & \nodata  &   0.026 &   0.226 &   0.248 &   0.024 &   0.161 &   0.143 &   0.217 &   0.250 &   0.238 &   0.237 &   0.208 &   1.716 &     9.4 &    20.2 \\
II~Zw~40 &   0.086 &   0.091 &   1.028 &   2.034 &   3.336 &   6.241 &   9.719 &  20.722 &  23.902 &  20.627 &  24.962 &  23.463 &  54.652 &   134.6 &    1616 &    4787 &    1427 \\
      &   0.005 &  \nodata   &   0.044 &   0.058 &   0.092 &   0.161 &   0.242 &   0.534 &   0.815 &   0.494 &   0.903 &   0.869 &   1.391 &     2.1 &      15 &      48 &     109 \\
Mk~5 &   0.308 &   0.410 &   0.946 &   1.550 &   1.902 &   2.984 &   3.198 &   3.509 &   3.802 &   2.742 & \nodata  &   6.745 & \nodata  &  19.041 &  16.118 &   236.8 &   144.7 \\
      &   0.004 &   0.003 &   0.044 &   0.033 &   0.070 &   0.259 &   0.027 &   0.239 &   0.210 &   0.202 & \nodata  &   0.462 & \nodata  &   0.790 &   1.459 &    10.0 &    17.0 \\
CG~10 &   0.141 &   0.153 & \nodata  &   0.281 &   0.433 &   0.458 & \nodata  &   0.422 &   0.504 &   0.355 & \nodata  & \nodata  & \nodata  & \nodata  & \nodata  & \nodata  & \nodata  \\
      &   0.003 &   0.002 & \nodata  &   0.014 &   0.035 &   0.038 & \nodata  &   0.019 &   0.033 &   0.033 & \nodata  & \nodata  & \nodata  & \nodata  & \nodata  & \nodata  & \nodata  \\
I~Zw~18 &   1.293 &   1.176 &   0.881 &   0.905 &   0.795 &   0.731 &   0.649 &   0.601 &   0.452 &   0.565 & \nodata  & \nodata  & \nodata  & \nodata  &   5.827 &  36.058 & \nodata  \\
      &   0.010 &   0.006 &   0.043 &   0.022 &   0.018 &   0.017 &   0.019 &   0.037 &   0.041 &   0.085 & \nodata  & \nodata  & \nodata  & \nodata  &   0.863 &   4.595 & \nodata  \\
Was~5 &   0.247 &   0.299 &   0.322 &   0.501 &   0.685 &   0.814 & \nodata  &   0.986 &   0.786 &   0.667 & \nodata  & \nodata  & \nodata  & \nodata  & \nodata  & \nodata  & \nodata  \\
      &   0.004 &   0.003 &   0.013 &   0.014 &   0.019 &   0.011 & \nodata  &   0.054 &   0.072 &   0.096 & \nodata  & \nodata  & \nodata  & \nodata  & \nodata  & \nodata  & \nodata  \\
Mk~36 &   1.590 &   1.574 &   1.615 &   1.893 &   1.808 &   2.432 & \nodata  &   2.643 &   2.312 &   1.937 & \nodata  &   1.165 & \nodata  &   1.645 &  26.197 &   368.0 &  51.638 \\
      &   0.012 &   0.007 &   0.062 &   0.052 &   0.048 &   0.031 & \nodata  &   0.122 &   0.143 &   0.148 & \nodata  &   0.182 & \nodata  &   0.224 &   1.950 &    15.9 &   9.159 \\
UM~439 &   1.482 &   1.779 &   2.188 &   3.251 &   3.449 &   3.628 & \nodata  &   4.752 &   4.357 &   2.858 & \nodata  & \nodata  & \nodata  & \nodata  & \nodata  & \nodata  & \nodata  \\
      &   0.013 &   0.006 &   0.103 &   0.096 &   0.099 &   0.043 & \nodata  &   0.149 &   0.133 &   0.155 & \nodata  & \nodata  & \nodata  & \nodata  & \nodata  & \nodata  & \nodata  \\
Mk~750 & \nodata  & \nodata  &   1.603 &   2.235 &   2.687 &   3.108 & \nodata  &   3.054 &   2.938 &   2.968 & \nodata  & \nodata  & \nodata  & \nodata  & \nodata  & \nodata  & \nodata  \\
      & \nodata  & \nodata  &   0.069 &   0.066 &   0.074 &   0.040 & \nodata  &   0.107 &   0.152 &   0.159 & \nodata  & \nodata  & \nodata  & \nodata  & \nodata  & \nodata  & \nodata  \\
UM~461 &   0.462 &   0.534 &   0.618 &   0.974 &   1.052 &   1.251 & \nodata  &   1.328 &   1.205 &   0.637 &   0.787 &   0.625 &   0.826 &   1.659 &  34.178 &   272.1 & \nodata  \\
      &   0.010 &   0.007 &   0.031 &   0.033 &   0.037 &   0.016 & \nodata  &   0.067 &   0.087 &   0.118 &   0.148 &   0.130 &   0.160 &   0.225 &   2.164 &    16.3 & \nodata  \\
UM~462 &   2.832 &   3.232 &   3.480 &   4.529 &   4.622 &   5.629 & \nodata  &   5.572 &   4.840 &   3.163 &   3.452 &   3.262 &   4.943 &   8.185 &   115.9 &   866.6 &   278.1 \\
      &   0.022 &   0.016 &   0.215 &   0.163 &   0.170 &   0.067 & \nodata  &   0.205 &   0.147 &   0.175 &   0.327 &   0.317 &   0.439 &   0.546 &     4.0 &    19.7 &    21.9 \\
Mk~67 &   0.237 &   0.371 &   0.767 &   1.064 &   1.293 &   1.565 &   1.800 &   2.268 &   1.988 &   1.247 & \nodata  & \nodata  & \nodata  & \nodata  & \nodata  & \nodata  & \nodata  \\
      &   0.004 &   0.003 &   0.035 &   0.027 &   0.032 &   0.039 &   0.051 &   0.094 &   0.099 &   0.113 & \nodata  & \nodata  & \nodata  & \nodata  & \nodata  & \nodata  & \nodata  \\
Mk~475 &   0.318 &   0.436 &   0.687 &   0.992 &   1.292 &   1.452 & \nodata  &   1.883 &   1.698 &   1.297 &   1.029 &   0.811 &   0.751 &   0.950 &  10.571 &   117.3 & \nodata  \\
      &   0.011 &   0.003 &   0.027 &   0.028 &   0.035 &   0.019 & \nodata  &   0.094 &   0.097 &   0.098 &   0.171 &   0.151 &   0.148 &   0.167 &   1.167 &     7.5 & \nodata  \\
Mk~900 &   0.715 &   1.101 & \nodata  &   4.613 &   8.318 &  10.745 &  12.394 &  16.384 &  17.475 &  12.707 &  10.060 &   6.775 &  10.556 &  19.652 & \nodata  &   417.9 &   380.6 \\
      &   0.005 &   0.005 & \nodata  &   0.081 &   0.261 &   0.485 &   0.069 &   0.498 &   0.515 &   0.421 &   0.567 &   0.464 &   0.586 &   0.798 & \nodata  &    14.4 &    25.4 \\
Mk~324 &   0.700 &   0.818 &   1.514 &   2.780 &   3.726 &   5.321 &   5.167 &   6.874 &   7.593 &   5.593 & \nodata  & \nodata  & \nodata  & \nodata  & \nodata  & \nodata  & \nodata  \\
      &   0.008 &   0.005 &   0.056 &   0.013 &   0.103 &   0.147 &   0.024 &   0.215 &   0.224 &   0.191 & \nodata  & \nodata  & \nodata  & \nodata  & \nodata  & \nodata  & \nodata  \\
Mk~328 &   0.324 &   0.435 &   1.213 &   2.542 &   3.938 &   6.457 &   6.843 &  10.765 &  12.463 &   8.938 & \nodata  & \nodata  & \nodata  & \nodata  & \nodata  & \nodata  & \nodata  \\
      &   0.004 &   0.003 &   0.045 &   0.028 &   0.109 &   0.178 &   0.032 &   0.317 &   0.230 &   0.181 & \nodata  & \nodata  & \nodata  & \nodata  & \nodata  & \nodata  & \nodata  \\

\enddata
\tablecomments{
Fluxes come from directly from sources listed in Table~\ref{bcdobs};
when multiple observations in the same filter exist, we use the one
with the lowest uncertainty. No
Galactic extinction corrections have been applied, and no minimum
error floors have been enforced to these fluxes. \textit{GALEX} fluxes without
pipeline-determined uncertainties are assigned errors of 5\%.
}
\end{deluxetable*}
\end{turnpage}


\section{Observational Data}\label{obs}

Our primary sample of 18 actively star-forming galaxies is the same as
that of 
\citet{janowiecki14}, and includes a variety of BCDs and related
galaxies. Some are canonical BCDs (e.g., I~Zw~18), while some have
smooth outer isophotes (similar to a dE) with a strong central
starburst (e.g., Mk~900). Some 
have offset starbursts (e.g., Mk~36, Mk~750), dual starburst
regions (UM~461, Mk~600), cometary shapes (Mk~5), or large numbers of
active star formation sites (e.g., UM~439, UM~462, UM~323). Our sample
is faint ($\langle$M$_\textrm{B} \rangle$=$-16$~mag),
blue ($0$$<$B$-$V$<$$0.5$), metal deficient ($\langle 12+$log$($O$/$H$) \rangle
$$\sim$$8$), and less than 50~Mpc away \citep[see Table~1 in][for
  more details]{janowiecki14}.

Our BCD sample is not comprehensive, but instead is
representative and its members span the range of 
parameter space that BCDs typically occupy. Samples of BCDs have been
defined in various ways, beginning from their
identification as extragalactic HII regions \citep{sargentsearle70},
continuing through the definitions of \citet{tm81} and
\citet{gildepaz03}. All of these samples commonly include
dwarf galaxies which are compact (or merely small) and intensely
forming stars. However, conclusions about the evolution of BCDs can
depend strongly on sample selection \citep[see also
  Section~5 of][]{janowiecki14}. Rather than creating or adopting a
definition of BCDs, we instead select a sample of BCDs and BCD-like
galaxies to study the extremes of dwarf galaxy evolution.

Toward this end,
we include the galaxies in the Local
Volume Legacy (LVL) survey as a comparison sample. LVL is a
volume-limited survey of 258  
galaxies within $11$~Mpc, which includes flux observations from the
ultraviolet to the far-infrared. 
The LVL sample is a
particularly good comparison sample as its volume limited nature
means that a majority of its galaxies have stellar masses below
$10^9M_\odot$, similar to our BCD sample \citep{dale09}. In fact, some
of the galaxies in the LVL sample are classified as BCDs, as discussed
in Section~\ref{bcd}.
In this analysis, we do not consider the 
environment of LVL galaxies (or BCDs), and treat each galaxy
individually; further work is needed to explore the effects of the
environment on star formation processes in dwarf galaxies.


\subsection{Photometric observations of the BCD sample}
 
The multi-wavelength photometric observations of our BCD sample come
from many sources. Table~\ref{bcdobs} shows a complete summary of the
photometry for each object, and Table~\ref{bcddata} contains all of
the observed fluxes of our BCD sample. We also use gas-phase abundances from
spectroscopic observations of the BCDs from sources in the literature
\citep{zhao10, brinchmann08, izotov07}. These abundances allow us to
consider the chemical evolution of the BCDs and put them in context of
galaxy scaling relations.

In the UV, we use far-ultraviolet (FUV) and near-ultraviolet (NUV)
observations from the \textit{Galaxy Evolution Explorer}
\citep[\textit{GALEX},][]{martin05, 
  morrissey07}. Most of our galaxies were observed through our
Cycle~3 program (GI3-089), but six are from archival 
images. \textit{GALEX} acquires images in FUV ($1344-1786$\AA) with a $4.3''$
FWHM PSF, 
and in NUV ($1771-2831$\AA) with a $5.3''$ FWHM PSF. The images are
processed and calibrated with the standard pipeline and were
downloaded from the archive. The fluxes are measured in large
apertures on the calibrated images. These UV images are especially
sensitive to recent star formation, and provide a critical measurement
of the presence and impact of young stars in the BCDs.



Some of our optical photometry comes from
\citet{sudarsky95} and \citet{norton97}, which contain complete
details of their reduction and calibration. In brief, the UBVRI
optical CCD observations were carried out at the Kitt Peak
National Observatory\footnote{Kitt Peak National Observatory, National
  Optical  Astronomy Observatory, which is operated by the Association
  of Universities for Research in Astronomy (AURA) under cooperative
  agreement with the National Science Foundation.}
  (KPNO) 0.9m telescope in November/December~1989 and 
April~1990. The images were calibrated with observations of Landolt
standards, and total fluxes were measured. NIR JHK photometry was also
obtained at KPNO with IR arrays \citep{salzerelston92}. 
Table~\ref{bcdobs} shows which photometric measurements come from this
dataset (labeled ``SN''), and the fluxes are given in
Table~\ref{bcddata}.

We have expanded this existing optical/NIR photometry with new
observations from the WIYN\footnote{The WIYN Observatory is a joint
  facility of the University of Wisconsin-Madison, Indiana University,
  the University of Missouri, and the National Optical Astronomy
  Observatory.} 3.5-m telescope at KPNO. Complete details of
the observations, reductions, and calibrations are given in 
\citet{janowiecki14}. In short, the Minimosaic and OPTIC (Orthogonal Parallel
Transfer Imaging Camera) imagers were used to obtain optical
observations and the WHIRC \citep[WIYN High Resolution Infrared
  Camera,][]{meixner10} imager was used to obtain NIR
observations. Observations were taken between November~2008 and
April~2010, and were calibrated with catalog measurements from Sloan
Digital Sky Survey (SDSS) and Two Micron All Sky Survey (2MASS) of
stars in the fields with our targets. Table~\ref{bcdobs} 
shows which flux measurements come from our recent WIYN observations
(labeled ``W''), and our complete set of fluxes and uncertainties are
reported in Table~\ref{bcddata}.

We also have calibrated narrow band Ha observations of the BCD
sample which are used to determine SFRs from the standard prescription
of \citet{kennicutt98}.
%
%
%
These \Ha
fluxes were measured from narrow-band imaging
\citep{salzerelston92}, and the calibrated \Ha 
luminosities are given in Table~\ref{bcdha}, along with metallicity
values from the literature for the BCD sample. Distances in
Table~\ref{bcdha} use flow models (Virgo+GA+Shapley) from
the NASA Extragalactic Database 
(NED\footnote{This research has 
  made use of the NASA/IPAC Extragalactic 
  Database (NED) which is operated by the Jet Propulsion Laboratory,
  California Institute of Technology, under contract with the National
  Aeronautics and Space Administration.}),
and for I~Zw~18 we use the tip of the red giant branch
  distance from \citet{aloisi07}.

\begin{deluxetable}{cccc}
\centering
\tabletypesize{\normalsize} 
\tablecaption{\Ha observations and abundances of the BCDs
\label{bcdha}
}
\tablewidth{0pt}
\tablehead{
\colhead{Galaxy} & \colhead{\Ha luminosity} & \colhead{Distance} & \colhead{Z} \\
\colhead{} & \colhead{$\times 10^{39}$ [erg/s]} & \colhead{[Mpc]} & \colhead{$12 + $log$ ($O$/$H$)$}\\
\colhead{(1)} & \colhead{(2)} & \colhead{(3)} & \colhead{(4)}
}
\startdata
UM~323   & 38.5 & 25.6 & 7.96$^a$ \\ 
UM~408   & 35.2 & 47.5 & 7.74$^a$ \\ 
Mk~600   & 27.3 & 13.6 & 7.94$^a$ \\ 
II~Zw~40 & 40.3 & 11.1 & 8.09$^a$ \\ 
Mk~5     &  7.7 & 15.3 & 8.06$^a$ \\ 
CG~10    &\nodata&30.7 & \nodata \\
I~Zw~18  & 24.3 & 18.2 & 7.18$^a$ \\
Was~5    & 14.3 & 23.1 & 7.85$^b$ \\ 
Mk~36    &  6.8 &  8.4 & 7.82$^a$ \\ 
UM~439   & 16.2 & 15.9 & 8.08$^a$ \\ 
Mk~750   &  2.7 &  5.2 & 8.18$^b$ \\ 
UM~461   &  9.1 & 12.7 & 7.81$^c$ \\ 
UM~462   & 32.6 & 13.5 & 7.80$^c$ \\ 
Mk~67    & 17.4 & 18.7 & 8.08$^a$ \\ 
Mk~475   &  6.9 & 11.9 & 7.93$^a$ \\ 
Mk~900   & 84.0 & 18.9 & 8.07$^a$ \\ 
Mk~324   & 31.3 & 23.2 & 8.18$^d$ \\ 
Mk~328   & 23.8 & 20.6 & 8.64$^d$ \\ 
\enddata
\tablecomments{
Column~3: distances from NED.
Column~4: gas phase abundance $Z$~=~$12+$log$($O$/$H$)$. Sources of
abundances are: $^a$ $T_e$
  abundances from \citealt{zhao10}; $^b$ N$2$ abundance from
  \citealt{zhao10}; $^c$ \citealt{brinchmann08}; 
  $^d$ \citealt{izotov07};
}
\end{deluxetable}

\subsection{Observations of the LVL sample}


The multi-wavelength observational data of the 258 galaxies in 
the LVL sample are nearly identical in quality and wavelength
coverage to our observations of the BCD sample. The SEDs for the
complete LVL sample are presented in 
\citet{cook14b}. These SEDs include \textit{GALEX} FUV and NUV images
\citep{lee11}, ground-based optical UBVRI images \citep{cook14a}, NIR
JHK images from 2MASS \citep[Two Micron All Sky Survey][]{dale09},
and Spitzer IR images \citep{dale09}. This observational
dataset is used to determine the multi-wavelength SEDs from
$1500$\AA$-24\mu$m for galaxies in the LVL sample. Constructing SEDs
with self-consistent UV, optical, NIR, and FIR fluxes is a complex
process, and care was taken to measure the galaxies within a common
aperture across all wavelengths \citep{cook14b}. In all, fluxes are
reported in 14 band-passes between \textit{GALEX} FUV and Spitzer $24\mu$m for
all of the 258 LVL galaxies. As noted by \citet{cook14b}, there are 47
upper limits on non-detections in the IR observations, 13 upper limits
in the UV observations, and 1 upper limit in the optical data. In
total, 3551 fluxes are used in the LVL SEDs.
\citet{cook14b} present the LVL SEDs and derive physical
properties from them, including their star formation rates, stellar
masses, and internal extinction

\begin{figure}
\centering
\includegraphics[width=7.2cm]{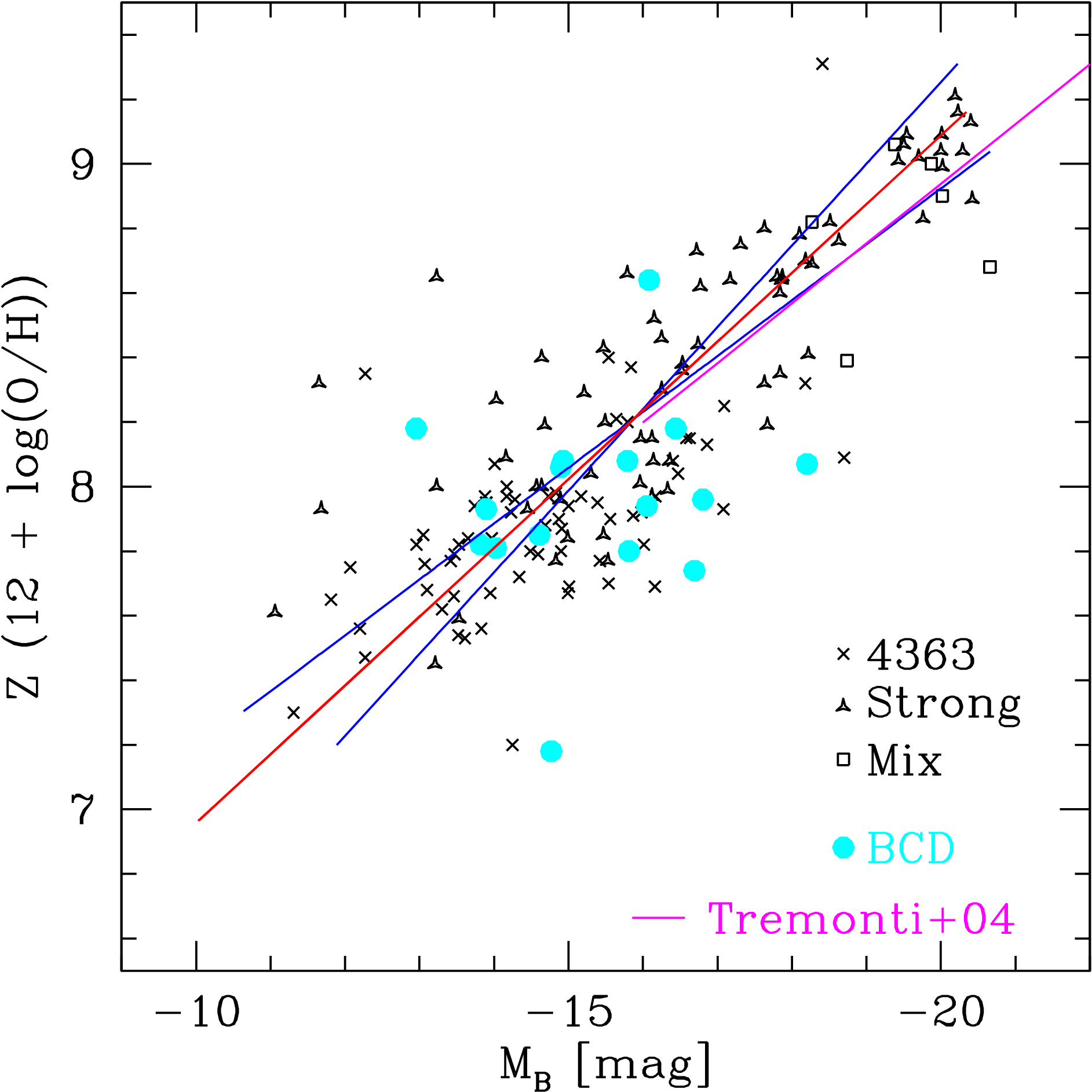}
\caption[L-Z relationship for LVL and BCD sample]{Different symbols
  show the L-Z  relation for all of the LVL 
  galaxies with measured gas-phase metallicities (various
  methods).
The solid blue lines show the best forward-fit and reverse-fit 
  to this relationship, and the red line shows their average.
Members of our BCD sample are shown as blue points at their
measured abundances. 
The magenta line shows the best-fit L-Z relation from
\citet{tremonti04}, which
agrees well with our fit. 
\label{MZ}}
\end{figure}



In addition to the complete set of panchromatic SEDs, gas-phase
metallicities have been measured for much of the LVL
sample. \citet{cook14b} compiles metallicity measurements from
\citet{marble10}, \citet{berg12}, and \citet{moustakas10}, which
come from different metallicity methods calibrations
\citep[see][]{marble10, mcgaugh91, pilyuginthuan05,
  kobulnickykewley04}. Of the 258 galaxies in LVL, 155 have measured 
metallicities (about half ``direct'' and half ``strong-line''
methods), which are shown in the luminosity-metallicity relationship
(L-Z) in Figure~\ref{MZ}.
Also shown in Figure~\ref{MZ} is the L-Z
relation from \citet{tremonti04} for a sample of $\sim$$53,000$
SDSS galaxies brighter than M$_\textrm{B}$=$-$$16$~mag, which agrees well with
the L-Z relation from the LVL and BCD samples.


\subsection{BCDs in LVL}
\label{bcd}

Some of the galaxies in LVL have been classified as BCDs -- in
particular Mk~475 is a member of both our BCD sample and of the LVL
sample. For brevity, throughout this work we refer to 
the ``BCD sample'' and the ``LVL sample'', but there exists a
continuum of galaxy properties across both samples. The LVL sample
\citep{dale09} includes both early- and 
late-type galaxies, and in particular includes 7~objects which
identified as BCDs in the Palomar/Las Campanas Atlas of BCDs
\citep{gildepaz03, kennicutt08}.

On subsequent plots and figures we take care to indicate these
individual populations (BCD, irregular/spiral, early-type) within the
LVL sample. Our goal is not to demarcate BCDs from ``normal'' dwarf
galaxies, but rather to explore the 
extremes of dwarf galaxy evolution. BCDs are often characterized by
their significant star formation rates, but they are unique in other
aspects as well. Throughout this work we will discuss the continuum of
dwarf galaxy properties, and show the extreme position that
BCDs occupy.


%
%
%


\begin{figure*}
\centering
%
%
%
\begin{overpic}[width=1.99\columnwidth]{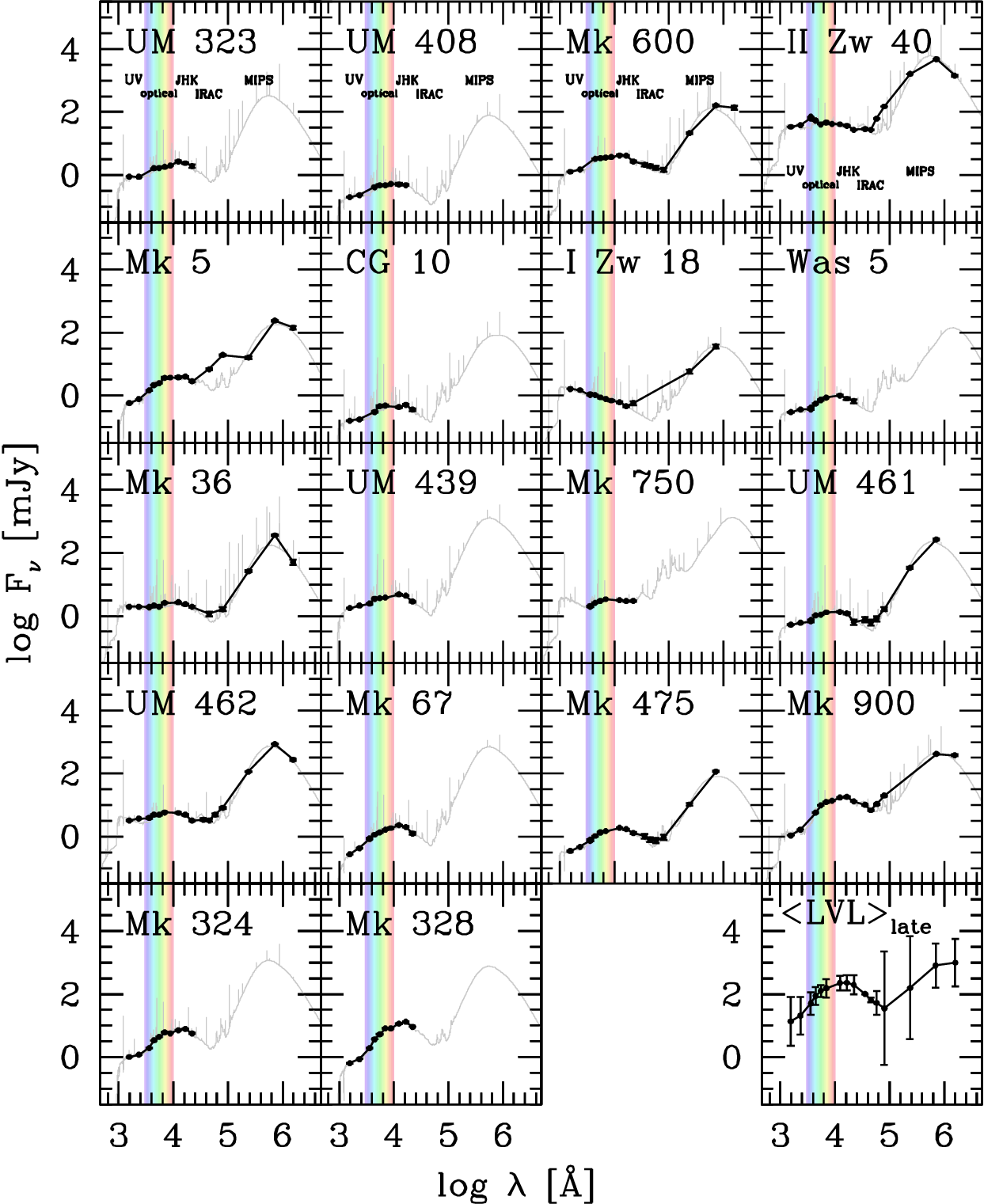}
 \put (20.91,81.6) {\includegraphics[width=1.1cm]{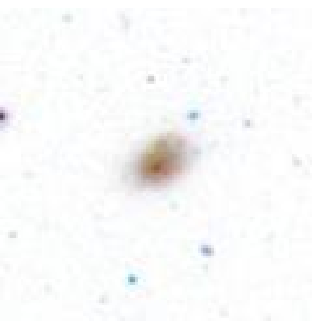}}
 \put (20.91,45.0) {\includegraphics[width=1.1cm]{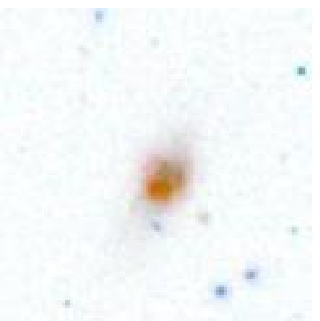}}
 \put (20.91,26.69) {\includegraphics[width=1.1cm]{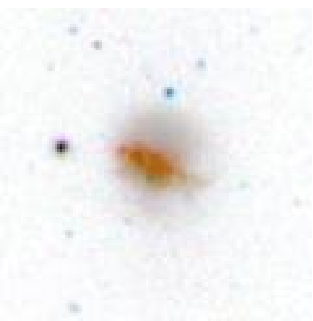}}
 \put (20.91, 8.32) {\includegraphics[width=1.1cm]{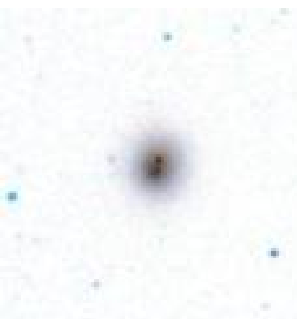}}

 \put (39.2,81.6) {\includegraphics[width=1.1cm]{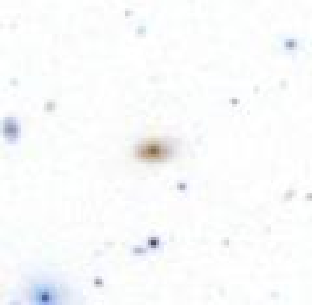}}
 \put (39.2,63.3) {\includegraphics[width=1.1cm]{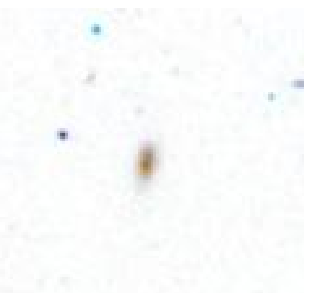}}
 \put (39.2,45.0) {\includegraphics[width=1.1cm]{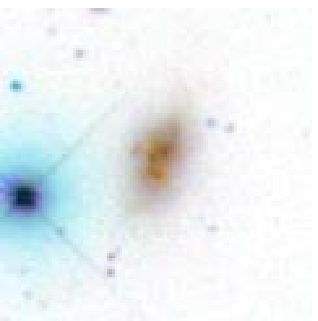}}
 \put (39.2,26.69) {\includegraphics[width=1.1cm]{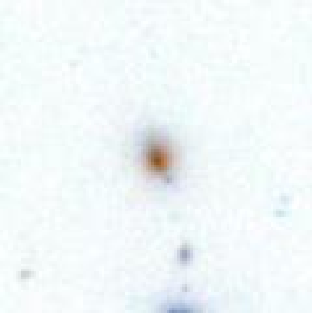}}
 \put (39.2, 8.32) {\includegraphics[width=1.1cm]{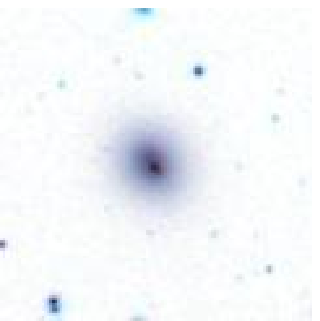}}

 \put (57.49,81.6) {\includegraphics[width=1.1cm]{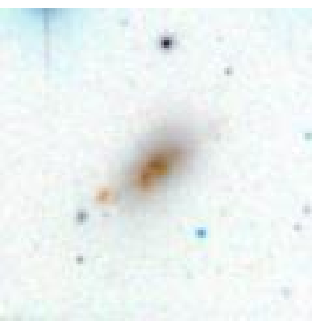}}
 \put (57.49,63.3) {\includegraphics[width=1.1cm]{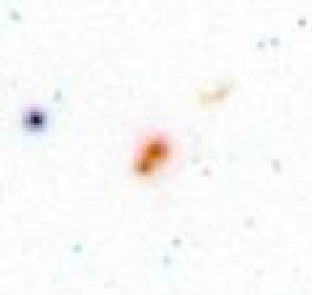}}
 \put (57.49,45.0) {\includegraphics[width=1.1cm]{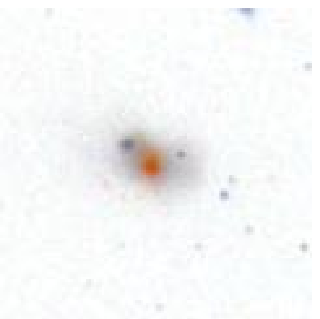}}
 \put (57.49,26.69) {\includegraphics[width=1.1cm]{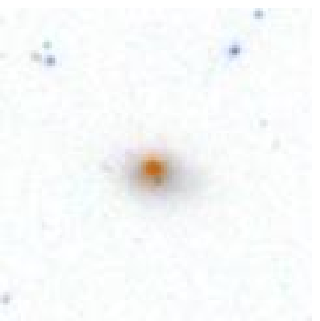}}

 \put (75.84,63.3) {\includegraphics[width=1.1cm]{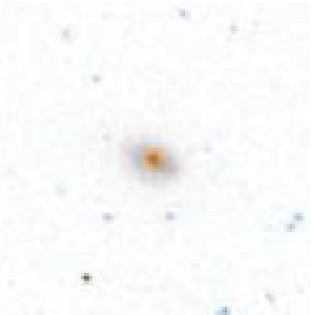}}
 \put (75.84,45.0) {\includegraphics[width=1.1cm]{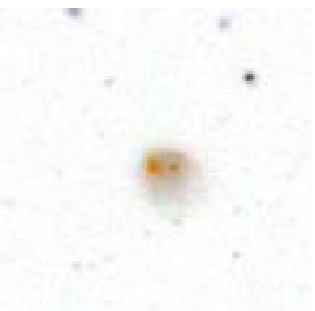}}
 \put (75.84,26.69) {\includegraphics[width=1.1cm]{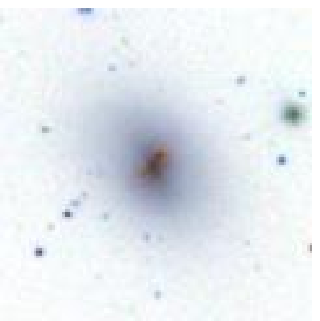}}
 \put (75.21,81.65) {\includegraphics[width=1.1cm]{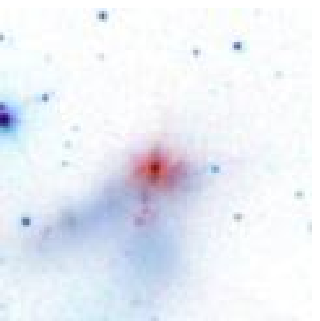}}

\end{overpic}
\caption[All SEDs for BCD sample]{All SEDs for our BCD samples. In
  each panel, the flux 
  density (in mJy) in each filter is shown
  by a dot at that filter's 
  effective wavelength. The photometric uncertainties are shown as
  error bars on the flux points, but are nearly always smaller than the
  points. Color thumbnails $100''$$\times$$100''$ from SDSS are shown
  in inverse video on
  each panel, where available. 
  An average SED for late-type low-mass LVL galaxies is shown in the 
  bottom right panel. It includes 92 LVL galaxies (none classified as
  BCDs) with stellar mass $<10^9$\msun, with measured fluxes in all
  filters, normalized to F$_{3.6\mu}=100$~mJy,
  and with error bars showing standard deviations at each wavelength.
  Light grey lines show the full spectra from the best-fitting
    SED models for each BCD. 
\label{allsed}}
\end{figure*}

%

\subsection{SEDs}

Figure~\ref{allsed} shows the complete set of observed SEDs for our
sample of BCDs. The SEDs have been corrected for Galactic extinction
\citep[via][for consistency with the LVL photometry]{sfd98} 
and are shown in flux units of mJy. Also shown are $100''$$\times$$100''$
color cutout images from SDSS Data Release 12
\citep[DR12,][]{dr12}, where available. The SEDs shown in
Figure~\ref{allsed} 
demonstrate the amount and quality of the photometric observations of
this sample. Most of the BCDs have UV fluxes, and many have complete
Spitzer FIR coverage as well. When compared with the average SEDs of
the low-mass late-type (non-BCD) LVL galaxies \citep[e.g., Figure~5 of][,
and shown in our Figure~\ref{allsed}]{cook14b}, the BCDs show 
much flatter (bluer) SEDs at UV/optical wavelengths. This is consistent with
the presence of substantial recent star formation activity. The three
most massive galaxies in our sample (Mk~324, Mk~328, 
Mk~475, Mk~900) have noticeably redder optical SEDs, and also have
smooth elliptical isophotes in their outskirts. These three galaxies
are classified as BCD type ``nE'' by \citet{gildepaz03}, and will be
referred to as ``BCD/E'' on subsequent plots.

Across the sample of BCDs, the UV slope shows significant variations from
the steep rise of Mk~67 to the flat slope of Mk~36. The UV slope is
sensitive to both the internal absorption from dust as well as the
current star formation. The IR observations are necessary to
disentangle the effects of dust and star formation. 

The detailed
shapes of these panchromatic SEDs encode much of the information
about the star formation history, stellar populations, and dust in the
BCD sample. The SEDs of each LVL galaxy are not reproduced here,
but are shown and discussed extensively in \citet{cook14b}.


\section{Fitting SEDs}\label{cigale}

We fit our SEDs with CIGALE \citep[Code Investigating Galaxy
Emission,][]{noll09}. CIGALE fits SEDs from UV to far infrared
in order to account for dust absorption and re-emission in a
self-consistent manner. It creates a grid of synthetic SEDs based on
theoretical models to account for all of the relevant line and
continuum emission and absorption from stars, gas, and dust. This grid
of models is then compared with observed SEDs in order to determine
the most likely values and uncertainties for various physical
parameters. In the following sub-sections we describe the models that
CIGALE uses to generate SEDs, our verifications of the appropriateness
of the models, and the consistency checks we employ to understand the
reliability of the fit results.

\subsection{Input Models}
\label{input_models}

CIGALE uses theoretical models to parametrize the flux emitted and
absorbed by the stars, gas, and dust in model galaxies, and produces
a grid of model spectra which are then converted to SEDs. 
\citet{noll09} describes the input models in complete detail,
and we briefly review each component and its contribution. 


The stellar emission is modeled using stellar population synthesis
models from \citet{bc03} with a Salpeter IMF, between metallicities of
$Z$$=$$0.0001$ and $Z$$=$$0.05$.
%
%
%
Two stellar populations are typically used: young and old stellar populations.
The light from the stellar
populations represents the dominant source of emission in 
the UV-optical range.

Nebular line and continuum emission are included in the UV-NIR range,
taking into account escape fraction and dust absorption. Using the
number of Lyman continuum photons (from the stellar continuum) to
compute the strength of the H$\beta$ line, a metallicity-dependent
template is used to determine the strengths of 119 other nebular
lines via an estimate of the number of ionizing photons
\citep{inoue11}.

Dust attenuation is handled with the formulas from
\citet{calzetti2000} and \citet{leitherer02}, based on the method of
\citet{ccm89}. 
In general, this requires computing and applying the attenuation curve 
(A$(\lambda)/$E$($B$-$V$)$) to all of the relevant flux-emitting components
considered in the model (both stellar and nebular emission). The
\citet{calzetti2000} attenuation law is used as a 
baseline but we allow its slope to change by multiplying it by
$(\lambda/\lambda_\textrm{V})^\delta$ where $\lambda_\textrm{V} = 5500$\AA, and 
$\delta$ ranges between $-1$ and $2.5$. 
This modification is required based on the variations seen in the
(limited number) of observations of the extinction law in other
galaxies \citep{witt00, inoue06}. The attenuation is combined with the
emission models as a
``negative'' flux, 
and is calculated separately for
each component of the model. The attenuation applied to the light 
coming from the old stellar population is reduced by a factor of
$f_\textrm{att} = 0.5$ from the value of the young population, to
account for the dustier nature of star forming regions. This has a
small effect. 

A key advantage of CIGALE's SED-fitting is its multi-wavelength energy
balance, and that consideration drives its treatment of infrared
re-emission from heated dust. CIGALE combines the amount of
attenuation present in the models with the semi-empirical re-emission
templates of \citet{dale14}. The templates are generated by
considering the contributions from a variety of dust heating
intensities and depend on a single heating parameter, 
$\alpha$. This exponent is the only free parameter in the dust
re-emission models as the total energy is constrained to be equivalent
to the amount that has been attenuated. 
%
%
%
%
%
CIGALE is also capable of modelling the emission from a 
dust-enshrouded AGN, but we do not include that option in our
fits; BCDs do not host AGN, and LVL contains mostly late-type dwarf
galaxies (NGC~855 is the most massive elliptical galaxy in sample, at
only $\sim$$10^9$\msun), so we do not expect any significant AGN emission.


\clearpage 
\begin{deluxetable*}{lccc}
\tabletypesize{\footnotesize}
\tablewidth{0pt}
\tablecaption{Input parameters for SED model grid
\label{grid}}
\tablehead{\colhead{Parameter} & \colhead{Symbol} & \colhead{Values} & \colhead{Units} }
\startdata
Old stellar population age               & age$_o$    & $1000, 3000, 10000$       & Myr  \\
Old stellar population $e$-folding time    & $\tau_o$   & 100, 1000, 10000         & Myr   \\
Young stellar population age             & age$_y$    & $30, 100, 300, 1000, 3000$  & Myr \\
Young stellar population $e$-folding time  & $\tau_y$   & $10, 100, 1000$           & Myr  \\
Young stellar population mass fraction        & $f_{b}$    & \multicolumn{1}{c}{$0.01, 0.03, 0.1, 0.3, 0.5$}  & \nodata \\
Stellar metallicity             & $Z$       &  \multicolumn{1}{c}{$0.0001,0.0004,0.004,0.008,0.02,0.05$}  & $Z/Z_\odot$ \\
Amount of dust attenuation      & E$($B$-$V$)_y$ & \multicolumn{1}{c}{$0.01, 0.02, 0.05, 0.1, 0.2, 0.3, 0.4, 0.5, 1$}  & mag \\
Power-law slope on extinction law & $\delta$ & \multicolumn{1}{c}{$-1, -0.5, 0, 0.5, 1, 1.5, 2, 2.5$}  & \nodata \\
Dust heating parameter          & $\alpha$ & \multicolumn{1}{c}{$0.5, 1, 1.5, 2.5$}  & \nodata \\
\enddata
\tablecomments{Note that we require 
 $\textrm{age}_o > \textrm{age}_y + 10$~Myr, to enforce a separation
  between the 
  old and young stellar populations.}
\end{deluxetable*}

CIGALE computes and applies the attenuation resulting from the effects
of the intergalactic medium (IGM), based on the prescription of
\citet{meiksin06}. While this attenuation is relatively small for the
nearby galaxies, we include it for consistency when comparing with
galaxies at different redshifts. The contribution of the IGM to the
overall attenuation is 
most relevant at very short wavelengths.

We can consider a wide variety of SFH scenarios in our models. CIGALE
currently includes three standard options: A double decreasing
exponential model, a ``delayed'' model, and a manually created SFH
supplied by the user. We primarily consider the double exponential
model, which describes the SFH as two exponential functions
characterized by $e$-folding time $\tau$. The age and $\tau$ of both
populations are free parameters, as is the mass fraction of the young
population compared to the old population. A minimum age separation
can be enforced between the 
two populations, and is typically set at $10$~Myr. In the following
sub-sections, we consider increasingly complex SFH assumptions and
the results of the SED fits.




\subsection{Preliminary SED-fitting tests}

To gain familiarity and confidence in the SED fitting methods and
results, we first considered simple SFH scenarios with very few free
parameters (e.g., a single recent burst on top of an old burst,
or on top of a constantly star-forming population). We varied grid
parameters, sampling densities, and explored the effects these choices
had on the outputs of the SED fits. Full details and results of these
extensive tests can be found in Chapter~3 of \citet{janowiecki15}. In
brief, we used a variety of diagnostics to check that the SED fits
were reasonable, and also to realistically assess the quality of the
outputs. In the following sub-sections, we discuss the grid choices
and quality assessments of the fits and their results.

\begin{figure*}[htb]
\includegraphics[width=5.65cm]{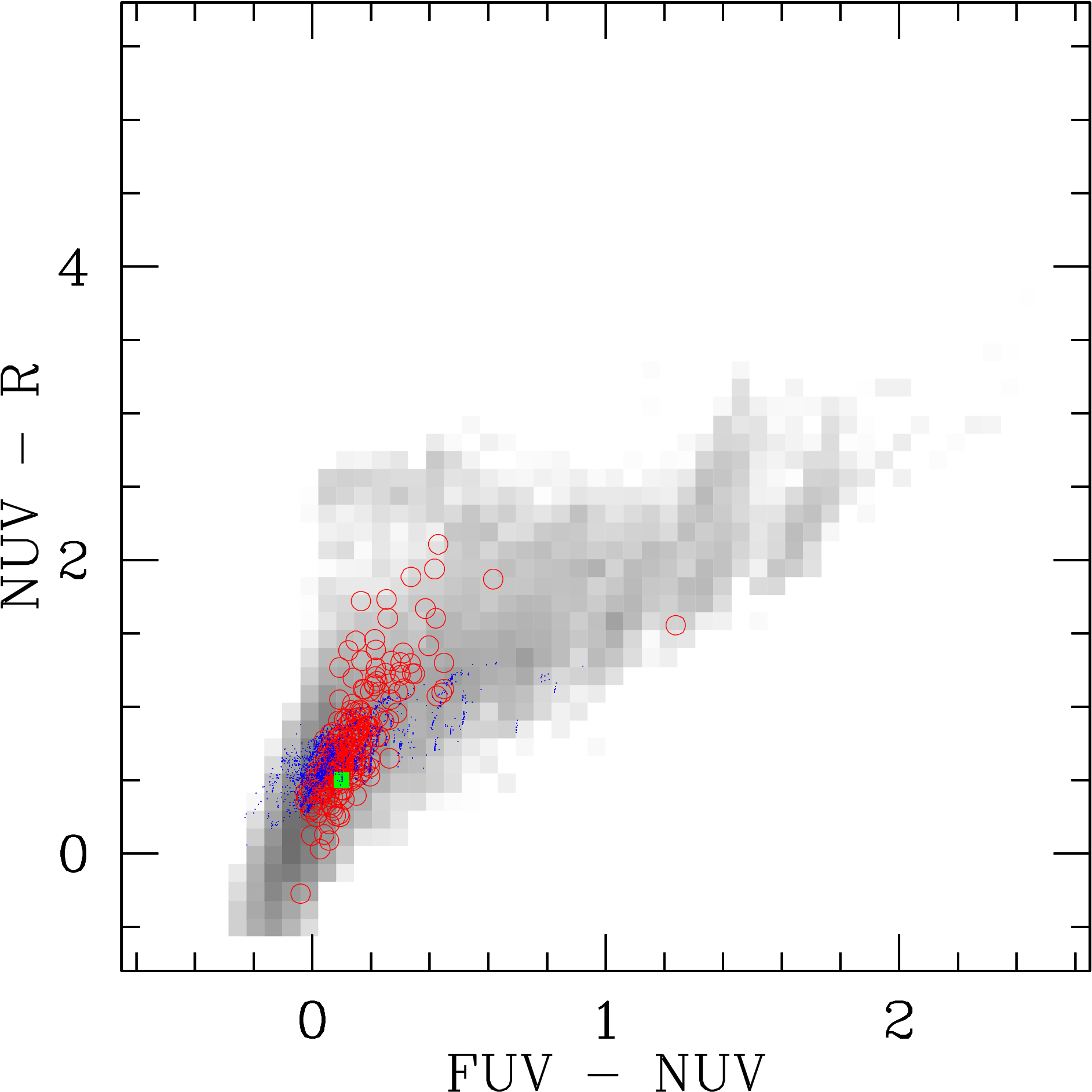} \hspace{0.25cm}
\includegraphics[width=5.65cm]{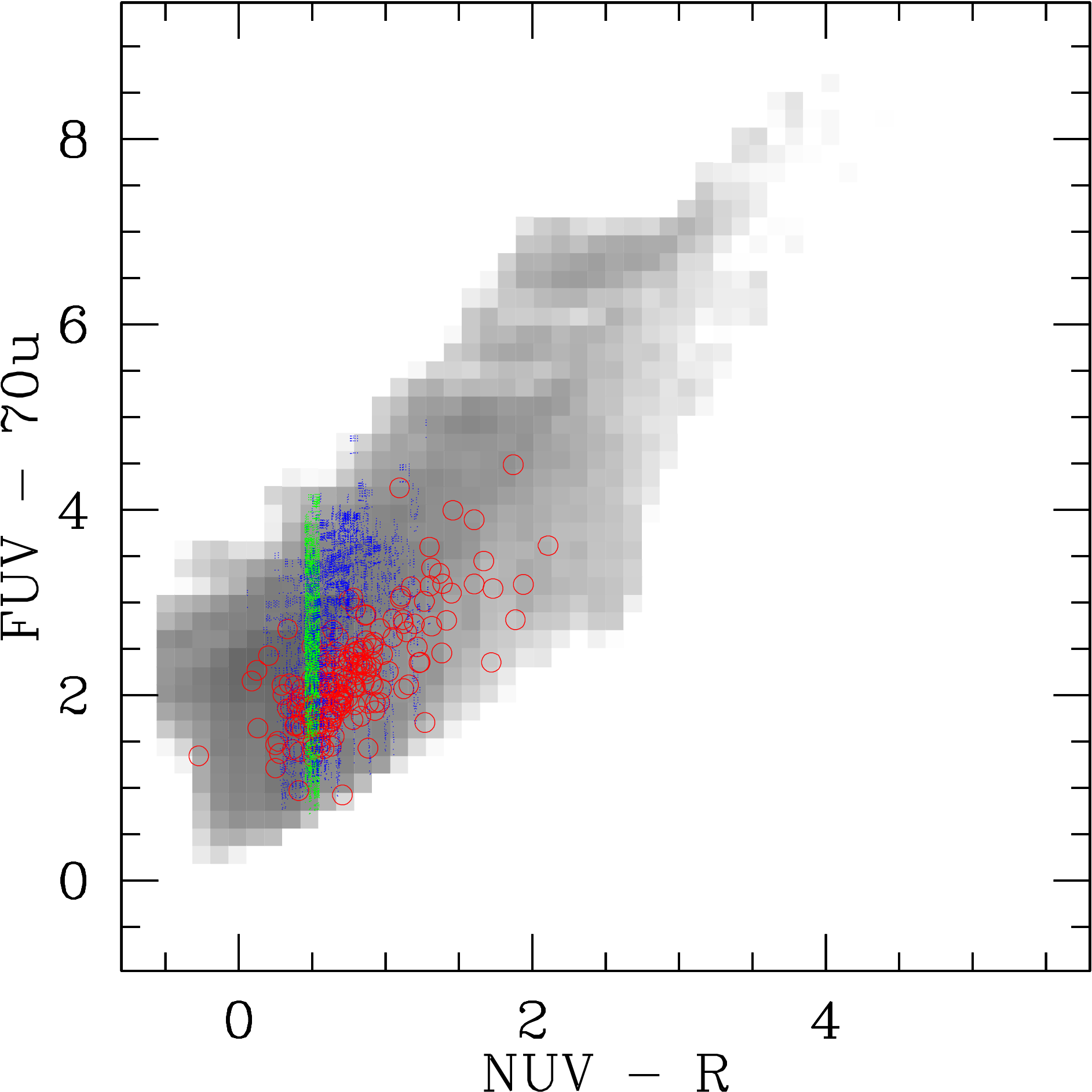} \hspace{0.25cm}
\includegraphics[width=5.65cm]{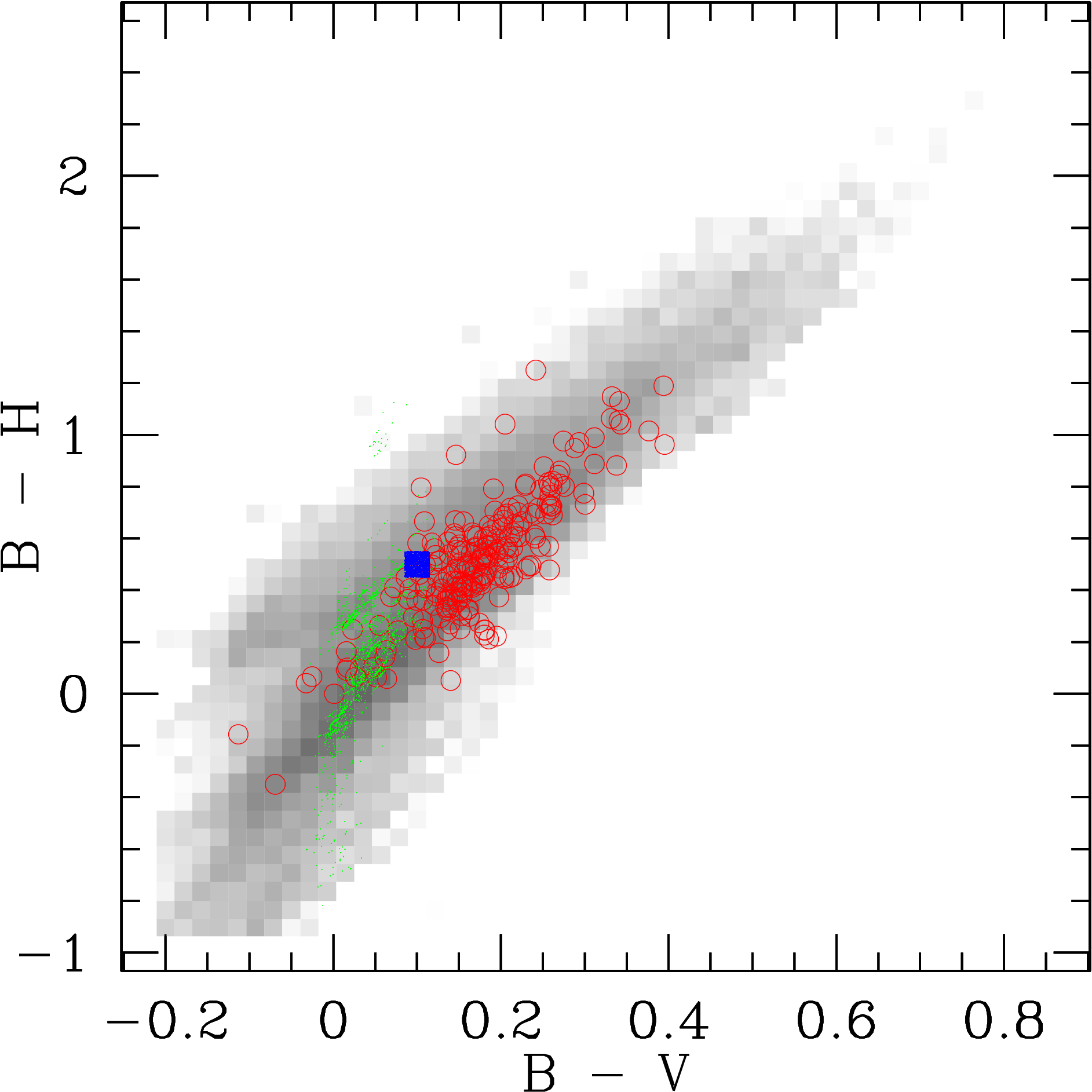}
\caption[Color-color diagrams]{Shaded grey cells
  indicate the number density of grid points 
  in that cell. Red circles indicate the observed colors of our BCD
  sample and LVL galaxies. Small green points show grid points within
  a single pixel bin on the left panel and their corresponding
  colors on the other panels. Small blue points are selected in a
  single pixel bin from right panel.
\label{cc}} 
\end{figure*}

\subsection{SFH and grid parameters}

In order to characterize a stellar population a SFH must be
assumed. We use CIGALE's double-exponential option to create a
two-population SFH. Both the old and young stellar populations are
described by declining SFRs which began abruptly and exponentially
decline. In these fits, both the age and exponential scale time are
allowed to vary, with the sampling of possible values shown in
Table~\ref{grid}. Note
that while the old stellar population age is allowed to be as young as
$1000$~Myr and the young stellar population is allowed to be as old as
$3000$~Myr, we require $\textrm{age}_o > \textrm{age}_y + 10$~Myr, to keep
a meaningful distinction between the old and young stellar
populations.

In this grid, the mass fraction of the young population is
allowed to vary between $1\%$ and $50\%$. The dust attenuation,
E$($B$-$V$)$, 
varies from 0.01 to 1.5 magnitudes, the Calzetti law is modified by a
power law slope between $-1$ and $2.5$, and the dust heating parameter
$\alpha$ varies between $0.5$ and $2.5$. The stellar metallicity is a
free parameter and 
varies between $Z=0.0001$ and $0.05$.
See Section~\ref{input_models} for
further details on these parameters. 

This grid requires $\sim$$1,000,000$ SEDs 
 to be computed, and the
process of generating and fitting the SEDs takes 
 about an hour on a
laptop computer. Once the entire library of SEDs for
each grid point is generated, they can be compared with our observed
SEDs. Before carrying out the actual fitting process, we
first demonstrate that the model grids are suitably well-matched and
appropriate for fitting our observed SEDs.

Following the method described in Section~3.2 of \citet{buat11}, we
test whether the colors of the model grids sufficiently overlap with
our observed SEDs by placing 
them onto diagnostic color-color diagrams. We consider the FUV$-$NUV and
NUV$-$R colors because of their connections with dust attenuation and
star formation history. We also show the FUV$-$$70\mu$ color for its
very large wavelength baseline, and the B$-$V and B$-$H colors as they are
commonly used. Figure~\ref{cc} shows these diagnostic color-color
plots. The galaxy with the most extreme FUV-NUV color is KDG~061, a
tidal dwarf galaxy in the M81 group. Its \textit{GALEX} observations
are very deep (16,238s exposure time), but its FUV flux is very
weak. Its optical colors are typical of dSphs but it has significant
amounts of HI and \Ha emission indicating ongoing star formation
\citep{johnson97, croxall09}.

Figure~\ref{cc} also demonstrates the correlations between these
colors in the models. Models are selected from within a single shaded
pixel on the left panel (shown in green) and plotted on the other
two plots at their appropriate colors. Similarly, models are selected
from a single pixel on the right panel (in blue) and are shown on the
other two diagrams. These model points demonstrate the connections
between the different color-spaces, and the strengths of
multi-wavelength SEDs from UV to IR.

Overall, it is clear
that our model grids are sampling an adequate amount of color-color
space to match most of the observations. Our choice of parameters for
the stellar populations and dust attenuation appear to cover an
appropriate range of values to be useful in fitting our
observations.

\subsection{Fitting SEDs and determining output values}

Now that we have shown that the colors of our model grids are suitable
comparisons to our observations, we can proceed to fit the SEDs.
To determine the best-fit values of each parameter for each observed
SED, CIGALE first calculates the $\chi^2$ value between the
observations and each model grid point SED. As described by
Equation~(5) 
from \citet{noll09}, this is calculated using:
\begin{equation}
\chi^2 (M_\textrm{gal}) = \Sigma^k_{i=1} 
\frac{(M_\textrm{gal} f_\textrm{mod,$i$} - f_\textrm{obs,$i$})^2}
{\sigma^2_{\textrm{obs,$i$}}}
\end{equation}
where the difference between each flux measurement
  ($f_\textrm{obs,$i$}$) and each model flux point ($f_\textrm{mod,$i$}$) is
  divided by the uncertainty on the observed flux
  ($\sigma^2_\textrm{obs,$i$}$), in filter $i$ \citep[see
    also][]{salim07}. The model fluxes are given per unit
  M$_\odot$, so are multiplied by the galaxy mass
  (M$_\textrm{gal}$). 
This summation
is taken over all flux observations ($k$) in the SED. The photometric
uncertainties on the flux 
observations, $\sigma_{\textrm{obs}}$, are included as a weighting
factor. 

After determining values of $\chi^2$ between a galaxy's SED and all of
the model grid points, CIGALE generates probability distribution
functions (PDFs) for selected parameters in a Bayesian-like framework
\citep{kauffmann03, salim05, salim07, noll09}. 
For each parameter, CIGALE creates a number of bins between its lowest
to highest values, and determines which models fall into each bin for
that parameter. Among the models in each bin, the model with
the highest probability (i.e., best match to observed SED) is found
and reported. With these maximum probabilities from each bin, a 
PDF can be generated which represents the maximum envelope of the
probability distribution for that parameter. These PDFs are used to
generate expectation values and uncertainties (see Figures~6 and 7 in
\citet{noll09} for further details about this method). 
The key analyzed parameters from the fits are described in
Table~\ref{parms}, their error determinations are discussed
further in Section~\ref{uncert}, and the best-fit values for the BCD
sample are shown in Table~\ref{best1}.

\begin{deluxetable*}{clc}
\centering
\tabletypesize{\scriptsize}
\tablecaption{Key analyzed parameters from SED fits
\label{parms}
}
\tablecomments{All input parameters are also analyzed.}
\tablewidth{12cm}
\tablehead{
\colhead{Parameter} & \colhead{Description} & \colhead{units}}
\startdata
$\langle$SFR$\rangle$$_{10}$   & SFR averaged over 10~Myr   & $M_\odot/$yr \\
$\langle$SFR$\rangle$$_{50}$   & SFR averaged over 50~Myr   & $M_\odot/$yr \\
$\langle$SFR$\rangle$$_{100}$  & SFR averaged over 100~Myr  & $M_\odot/$yr \\
$\langle$SFR$\rangle$$_{500}$  & SFR averaged over 500~Myr  & $M_\odot/$yr \\
$\langle$SFR$\rangle$$_{1000}$ & SFR averaged over 1000~Myr & $M_\odot/$yr \\
$\langle$SFR$\rangle$$_\textrm{all}$ & SFR averaged over lifetime       & $M_\odot/$yr \\
\hline
$M_{\star,o}$ & Stellar mass of old stellar population & $M_\odot$ \\
$M_{\star,y}$ & Stellar mass of young stellar population & $M_\odot$ \\
\enddata
\end{deluxetable*}

\begin{figure}[htb]
\centering
\includegraphics[width=7.5cm]{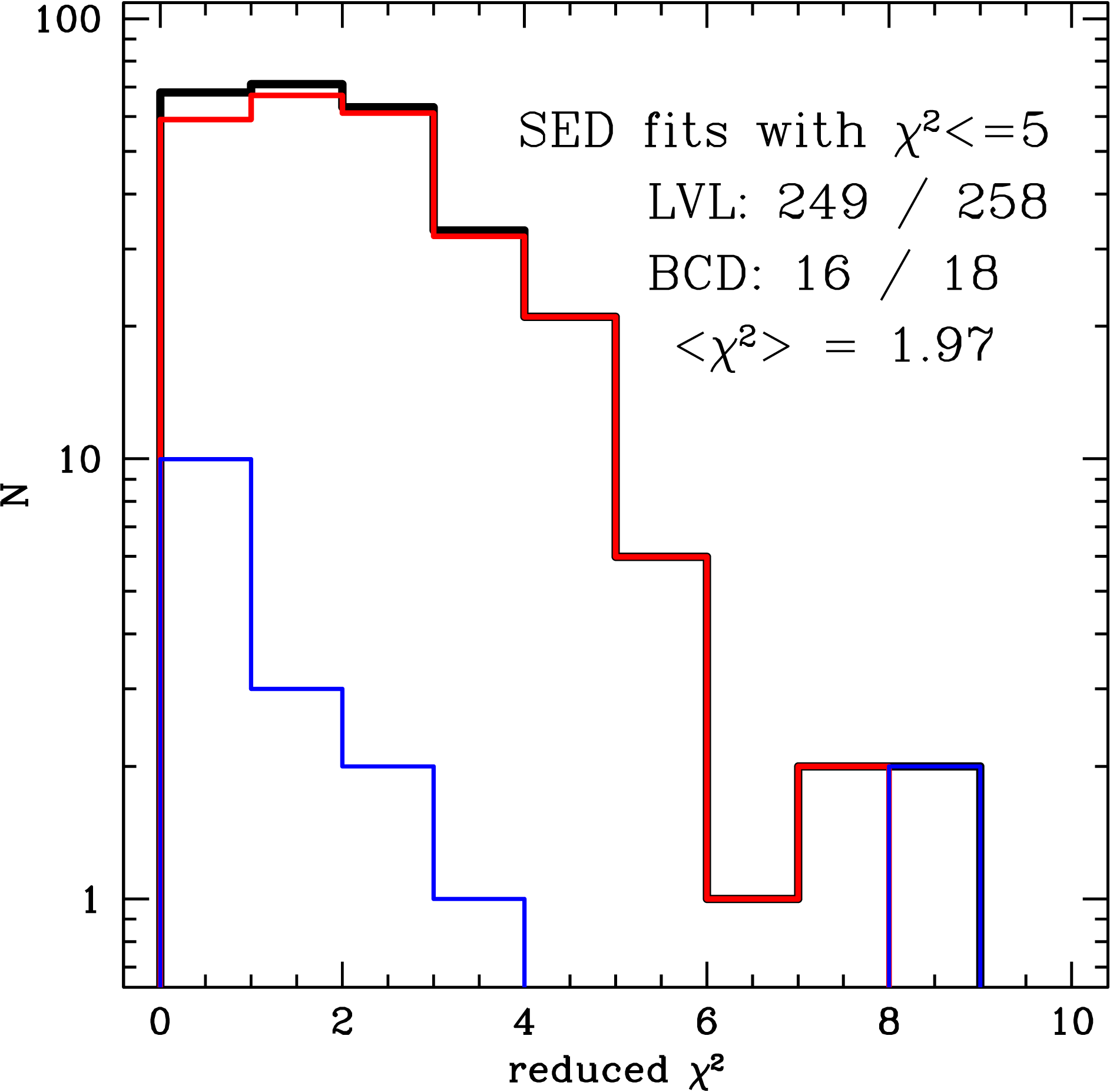}
\caption[Reduced $\chi^2$ distribution for our fits]{Distributions of
  values of reduced $\chi^2$ from our SED 
  fits. The thick black line shows the
  distribution of the combined LVL and BCD samples, while the BCDs are
  shown separately in blue and the LVL galaxies are shown separately
  in red. 
\label{x2}} 
\end{figure}

We use CIGALE to fit SEDs using all of the
fluxes from UV (FUV/NUV), optical (UBVRI), near-infrared (JHK), and
infrared (Spitzer IRAC-MIPS).
We examine the distribution of reduced $\chi^2$ values from the
best-fitting SEDs
to assess the success at fitting our
observations. Figure~\ref{x2} shows the reduced $\chi^2$ distribution
for our fits. On average, the reduced $\chi^2$ is $1.97$, and only
$11$ galaxies have $\chi^2>5$.

\subsection{Reliability and uncertainty estimates}
\label{uncert}

We now describe a few verifications and consistency checks which were
used to estimate the reliability and uncertainty of the SED fits.

\begin{figure*}[htb]
\centering
\includegraphics[width=\textwidth]{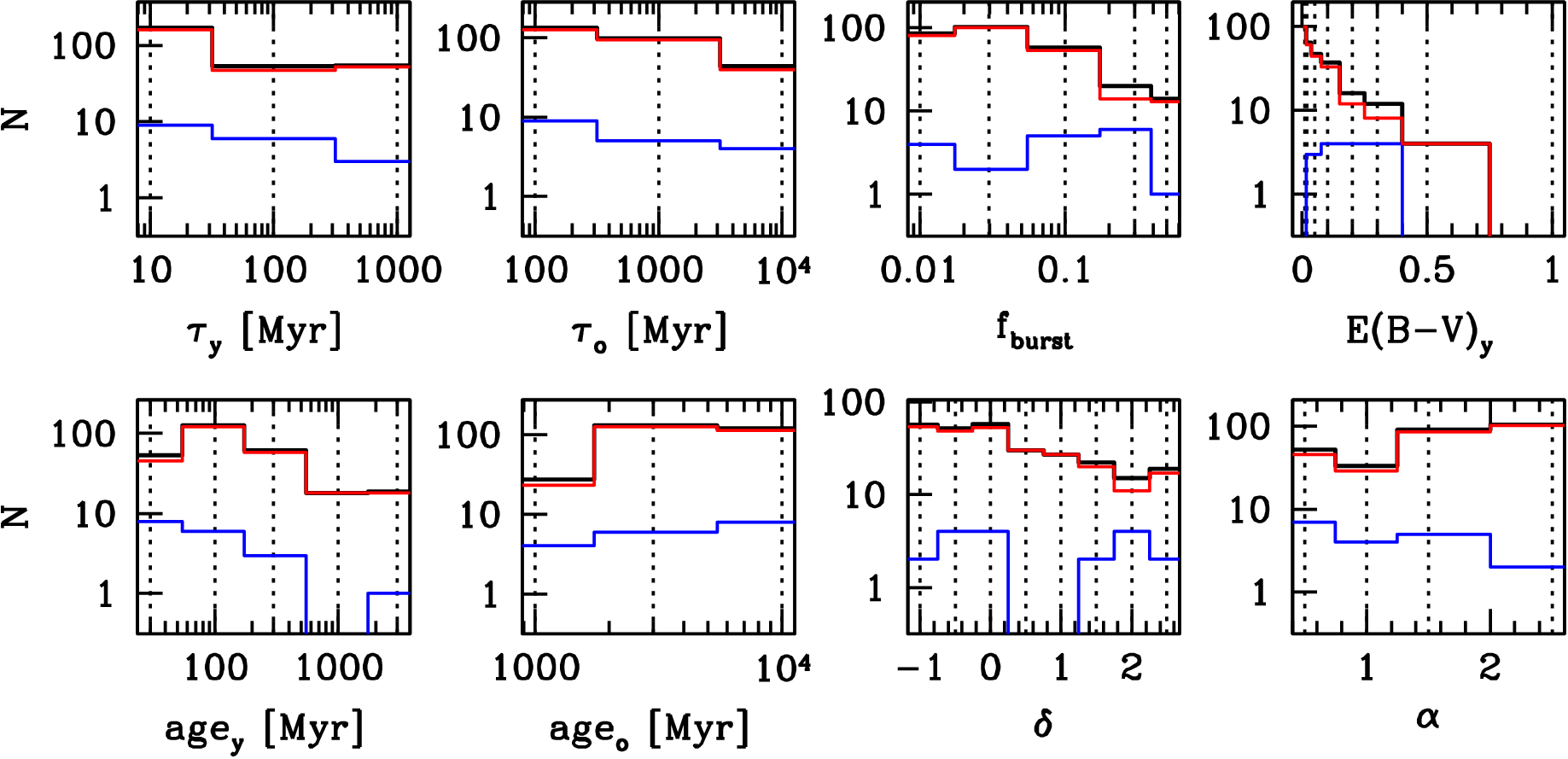}
\caption[Sample of parameter distributions]{Sample of
  parameter distributions. 
Vertical dotted lines show grid points.
Histograms of the best-fit results are shown in black for the complete
sample, in red for the LVL sample, and in blue for the BCD sample.
\label{hist1}}
\end{figure*}

It is important to verify that the parameter space covered by the grid
is of suitable resolution and range. Inadequate resolution
in a particular parameter will decrease the accuracy of that
parameter's determination, while over-sampling a poorly constrained
parameter can artificially increase its uncertainty or negatively
impact the determinations of other parameters
\citep{buat11}. Figure~\ref{hist1} shows histograms of the
best-fitting results for the most relevant parameters.

We experimented with different ranges and resolution of these
parameters until satisfactory distributions were obtained. Initially,
our grid was too narrow and some of the histograms showed
unrealistically narrow spikes at the 
extremes of parameter space. We 
expanded the range of parameter space (by expanding the maximum or
minimum values of the parameters) until there were no unrealistic
peaks at the edges.
Many of the parameters are logarithmically sampled in order to
smoothly cover the wide range of parameter 
space. The resolution in each parameter was also adjusted until a
generally smooth and continuous histogram was obtained, to eliminate
unrealistically sharp features which were artifacts of inadequate
sampling. The practical constraints of computing time were
also considered, which prevented the grids from becoming unmanageably
large. The resulting grid represents a compromise between
covering the necessary parameter space with enough resolution and
restricting the computational needs within reason. The final choices
for parameter sampling are shown in Table~\ref{grid}.

\begin{figure*}
\centering
\includegraphics[width=13.5cm]{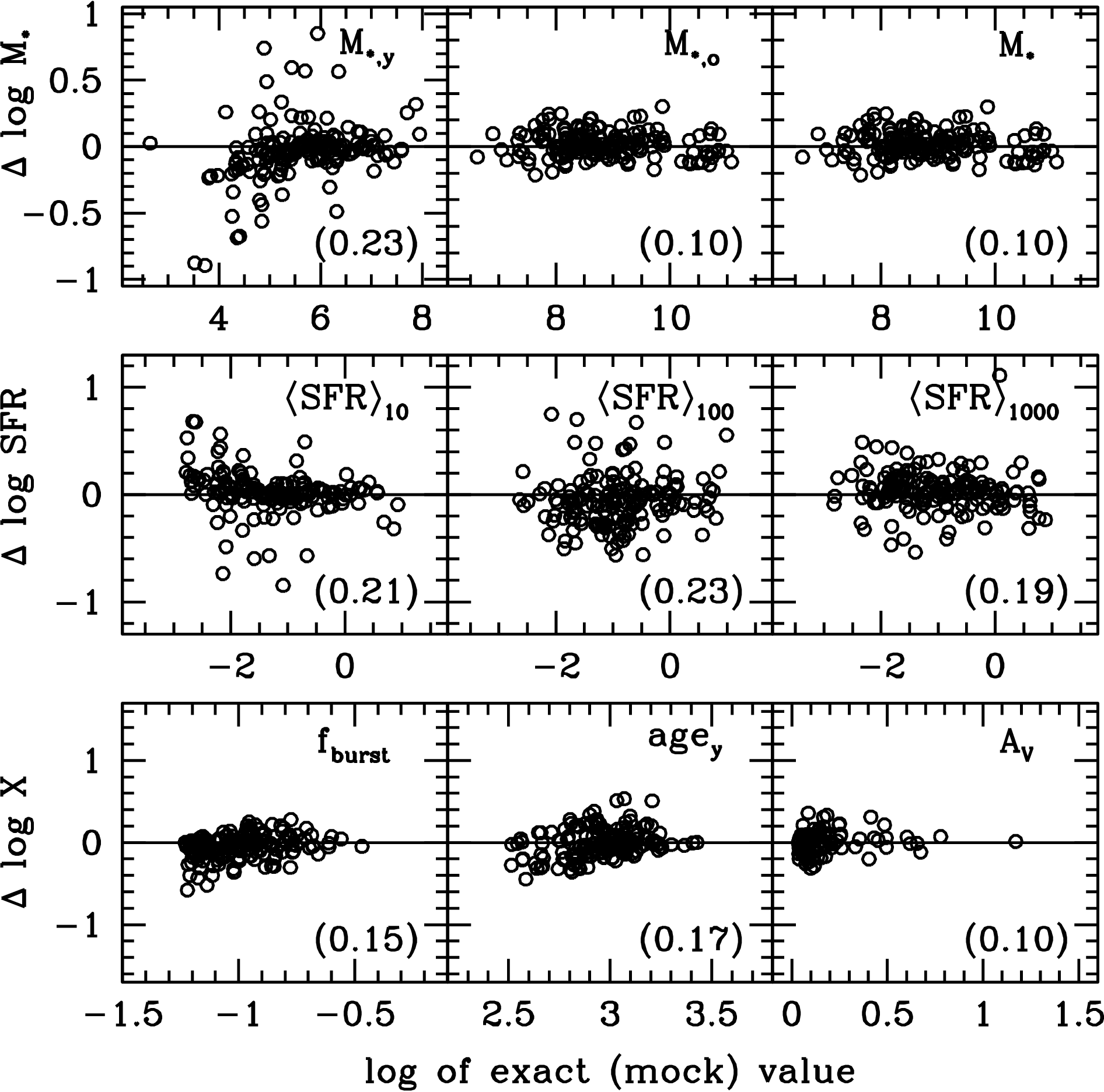}
\caption[Mock results]{Results from mock analysis for the combined
  BCD and LVL samples. 
For each of these nine selected parameters, the x-axis shows the exact
(``mock'') value while the y-axis shows the difference between the
exact and re-fitted value.
The top row shows the mass estimates for the young, old, and total
stellar populations, in units of solar masses.
The middle row shows three SFR estimates, averaged over 10, 100, and
1000~Myr, all in units of solar masses per year.
The bottom row shows the logarithmic differences for ``$X$'', where
$X$ in the first column is the mass fraction of the 
burst population, in the second column is the
age of the burst population (in Myr), and in the third column is the
dust attenuation (A$_\textrm{V}$, in magnitudes).
In all panels, the number in parentheses indicates the standard
deviation of the distribution of points shown.
\label{mock}} 
\end{figure*}

In order to estimate the reliability of the SED fit results for each
galaxy, we start with the best-fitting grid point for each
object. This best-fitting model SED is treated as a ``mock'' observation
(retaining the original 
photometric uncertainties on the real observations), and re-fitted with
the same grid to re-generate the most likely value of each
parameter. This ``mock'' fitting method allows us 
to estimate the reliability of the SED fit results \citep{salim09,
  giovannoli10} by comparing ``known'' input parameters of
the model SED with its re-fitted parameters.

Figure~\ref{mock} shows shows the results of our mock analysis for
nine of the analyzed parameters. As before, we only show the SED 
fits which have a reduced $\chi^2<5$ and which had a complete set of
observations for all $15$ flux 
points. Shown on the x-axes are the exact
values of each parameter used to generate the ``mock'' SEDs. The
y-axes show the differences between the exact ``mock'' values and the
best-fit values from the SED re-fitting process. The number in
parentheses at the bottom right corner of all panels is the rms
scatter for that parameter.

In some cases, the exact parameter values in the ``mock'' SEDs can be
reliability recovered by our fits. For example, the mass of the old
stellar population ($M_\textrm{o,stellar}$) shows no systematic trends
and a scatter of only $0.10$~dex. However, the mass of the young
stellar population ($M_\textrm{y,stellar}$) has more scatter and perhaps
a slope or offset at the lowest stellar masses. Furthermore, the
best-fit values of the age 
of the young stellar population (age$_\textrm{y}$) show a small
systematic trend where younger ages are likely to be under-estimated
and older ages are likely to be over-estimated.

\begin{figure*}[htb]
\centering
\includegraphics[width=0.47\textwidth]{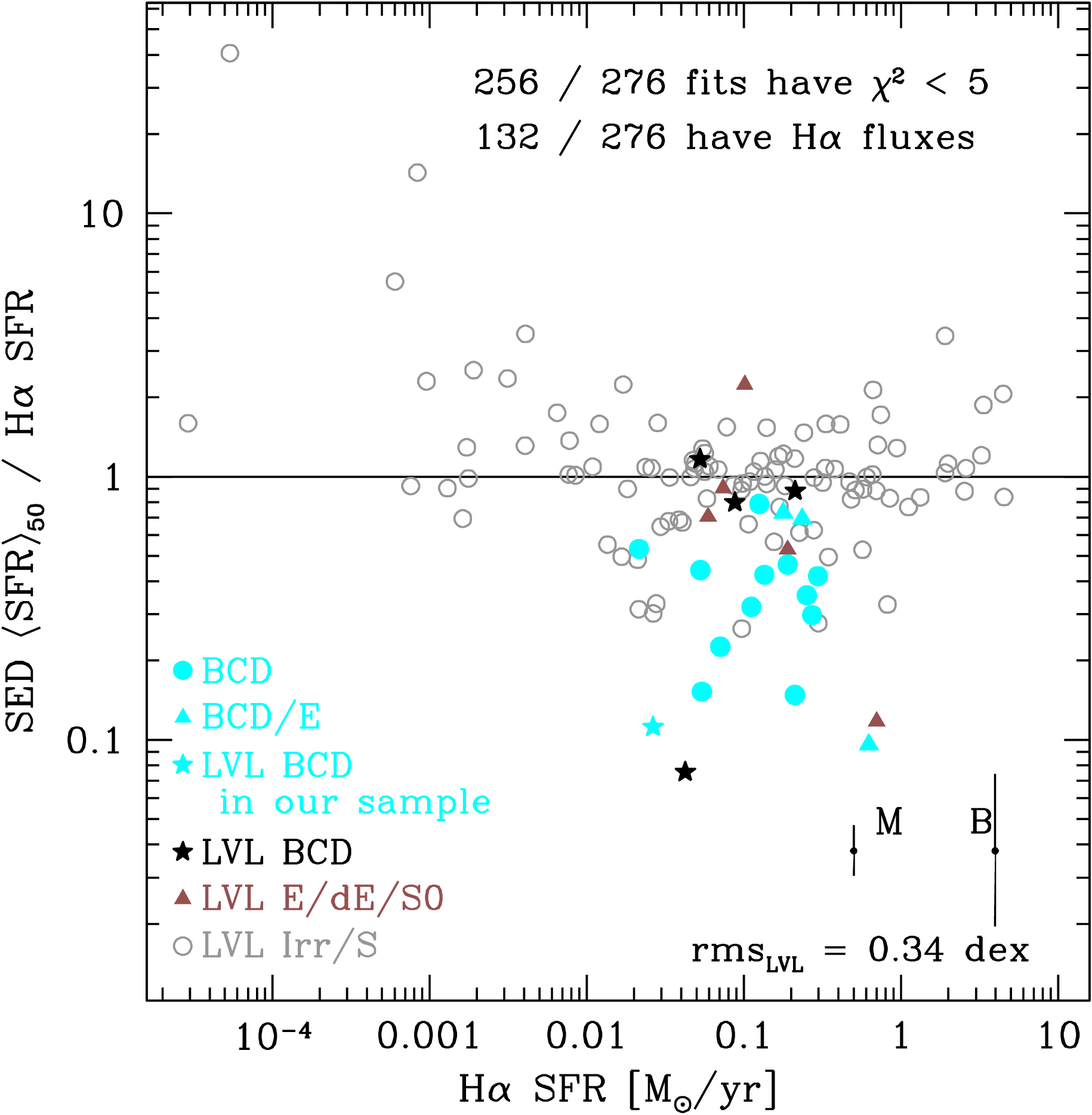}
\includegraphics[width=0.47\textwidth]{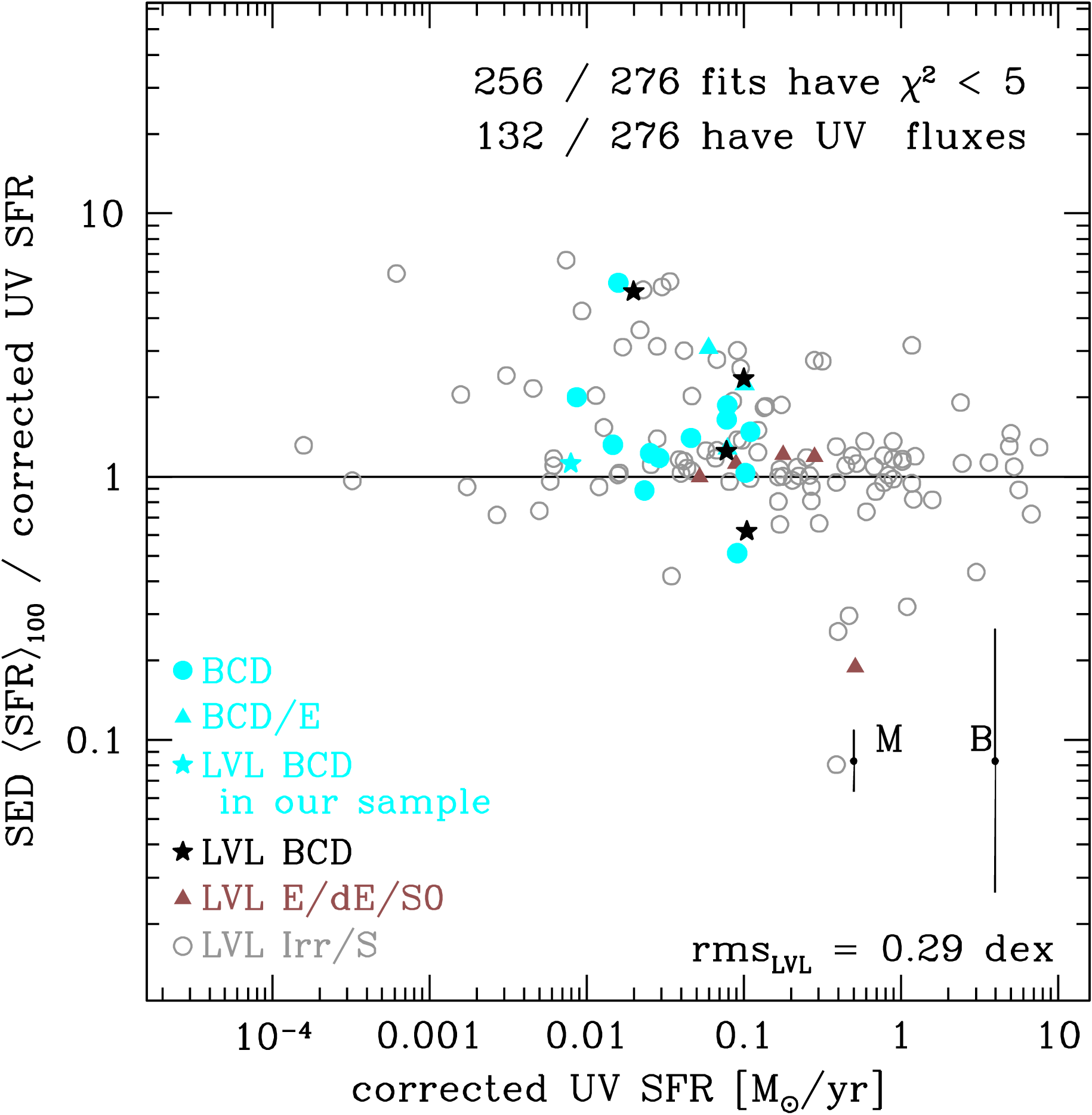}
\caption[SFR comparisons]{SFR comparisons between SED fits, \Ha, and
  UV observations. 
Galaxies are plotted by morphology; LVL 
    irregulars/spirals are grey circles, LVL E/dE/S0 are red triangles,
    LVL BCDs are black stars, and Mk~475 is shown as a cyan
    star. Our BCD sample is 
    plotted as cyan dots, with the nE BCDs shown as cyan triangles.
The left panel compares the 50-Myr average SED SFR
   and the \Ha SFR. 
The right panel compares the 100-Myr average SED
   SFR 
   and the SFR from the extinction-corrected UV luminosity (which is
   also included in the SED fit).
  The RMS scatter is measured for late-type LVL galaxies only.
The representative error bars at the bottom right of both panels show
   the average deviations from 
   the mock method (``M'') and the average Bayesian uncertainty
   (``B''). 
\label{sfr}} 
\end{figure*}

The reliability of these parameters varies in this mock analysis, but
these reliability estimates are only 
useful in conjunction with $\chi^2$ indicators and other independent
verifications of the SED fits.  The deviations shown here represent
the reliability of the fit results, independent from our actual
observations.
The differences in best-fit parameters of the mock SEDs and the
re-fit SEDs give an estimate of the reliability of this grid and
method, and show which parameters are more reliably determined in
this type of analysis. Some parameters are more reliably fit than
others (e.g., the stellar mass of the young population has a great
scatter than the stellar mass of the old population), which is
incorporated into the uncertainties listed in the error budget in
Table~\ref{err1}. This error budget also includes the full Bayesian
uncertainties generated from our fitting procedure.


\subsection{Further verifications}
\label{external}

\begin{figure*}[!htb]
\centering
\includegraphics[width=0.47\textwidth]{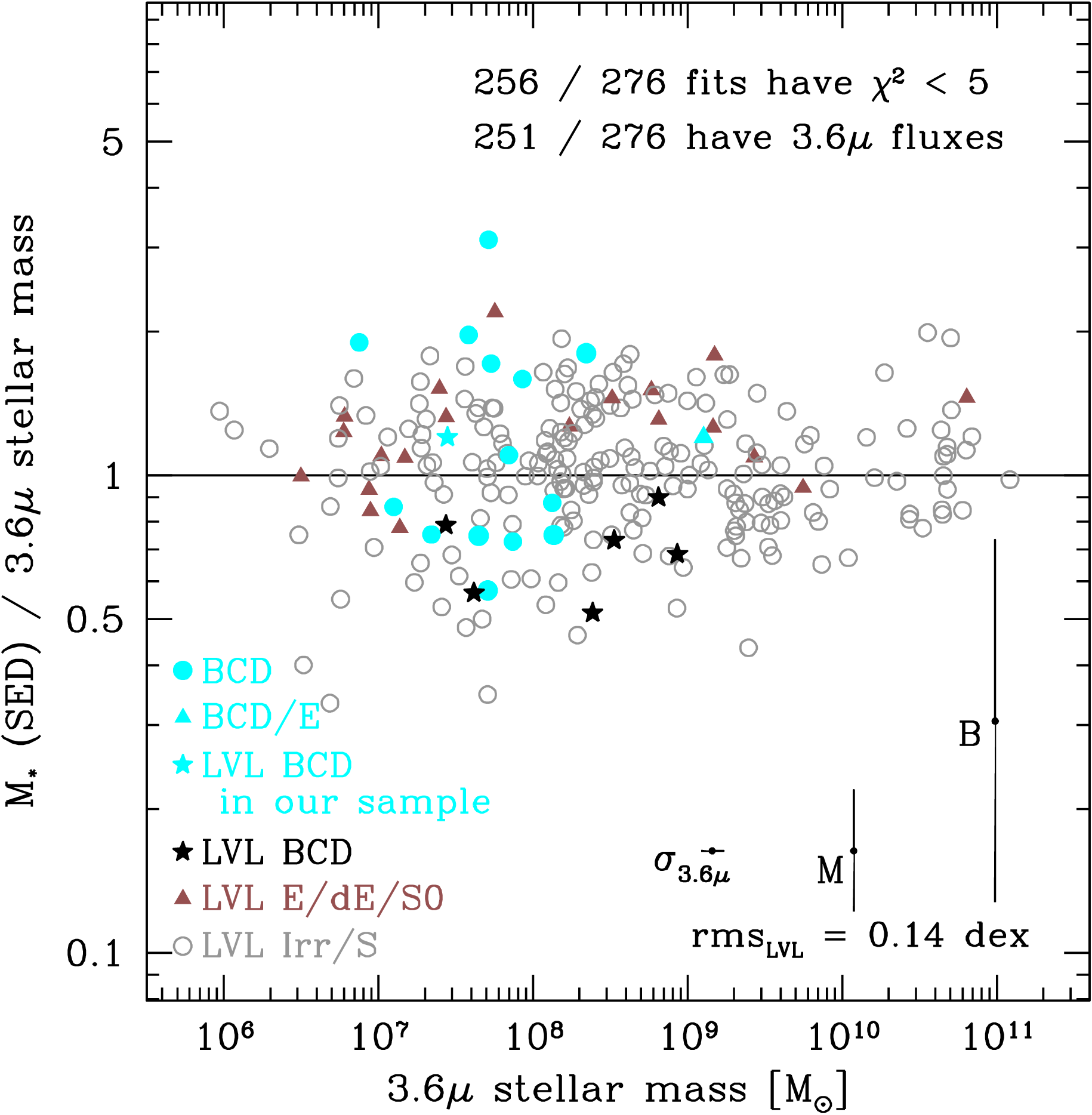}
\includegraphics[width=0.47\textwidth]{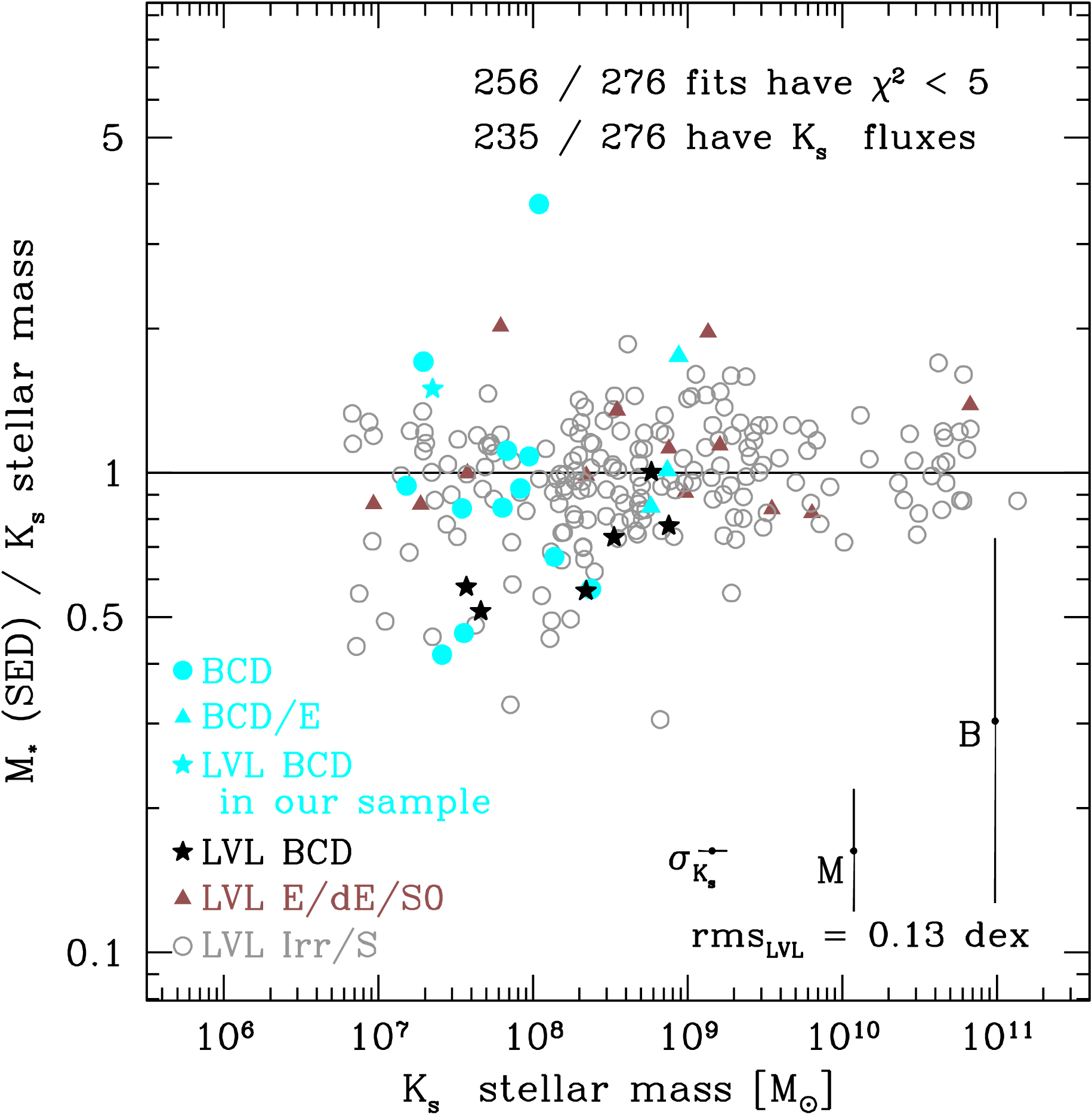}
\caption[Stellar mass comparisons]{Stellar mass
  comparisons between SED fits and simple M/L estimates, 
with the same color-coding as Figure~\ref{sfr}.
The left panel compares the stellar mass from the SED fits with that
    from the $3.6\mu$m luminosity, and the right panel compares with
    the stellar mass from K$_\textrm{s}$ luminosity. Both of these fluxes are
    themselves used in the SED fitting.
Even with the large uncertainty estimates from the mock and Bayesian
   methods, the agreement between $3.6\mu$m and K$_\textrm{S}$ luminosities and
   the SED-derived stellar masses is very good.
\label{M}} 
\end{figure*}

\hspace*{-2cm}
\begin{deluxetable*}{cccccccccccc}
\tabletypesize{\scriptsize}
\tablecaption{Best fit values and Bayesian uncertainties for derived parameters of BCDs
\label{best1}}
\tablewidth{0pt}
\tablehead{
\colhead{ Galaxy } &  
\colhead{ $N_\textrm{f}$ } &  
\colhead{ $\chi_\textrm{r}^2$ } &  
\colhead{ log M$_\textrm{$\star$,o}$ } &  
\colhead{ log M$_\textrm{$\star$,y}$ } &  
\colhead{ $\langle$SFR$\rangle_{50}$ } &  
\colhead{ $\tau_o$ } &  
\colhead{ $\tau_y$ } &  
\colhead{ age$_o$ } &  
\colhead{ age$_y$ } &  
\colhead{ $f_\textrm{burst}$ } &  
\colhead{ $A_V$ } \\ 
\colhead{} &  
\colhead{} &  
\colhead{} &  
\colhead{[$M_\odot$]} &  
\colhead{[$M_\odot$]} &  
\colhead{[$M_\odot$/yr]} &  
\colhead{[Gyr]} &  
\colhead{[Gyr]} &  
\colhead{[Gyr]} &  
\colhead{[Gyr]} &  
\colhead{} &  
\colhead{[mag]} \\ 
\colhead{(1)} &
\colhead{(2)} &
\colhead{(3)} &
\colhead{(4)} &
\colhead{(5)} &
\colhead{(6)} &
\colhead{(7)} &
\colhead{(8)} &
\colhead{(9)} &
\colhead{(10)} &
\colhead{(11)} &
\colhead{(12)}
}
\startdata
 UM323 &    9 & 0.3 & 8.13 (0.77) & 6.10 (0.71) & 0.124 (0.091) & 5.1 (3.3) & 0.42 (0.32) & 3.6 (2.3) & 0.76 (0.82) & 0.16 (0.15) & 0.32 (0.33) \\ 
 UM408 &    8 & 0.1 & 8.06 (1.56) & 5.91 (0.93) & 0.081 (0.089) & 4.9 (3.3) & 0.39 (0.31) & 3.4 (2.1) & 0.88 (0.90) & 0.18 (0.16) & 0.19 (0.25) \\ 
 Mk600 &   16 & 1.6 & 8.01 (0.37) & 5.50 (0.35) & 0.031 (0.010) & 5.1 (3.3) & 0.35 (0.29) & 3.7 (2.4) & 0.92 (0.85) & 0.19 (0.15) & 0.22 (0.03) \\ 
 IIZw40 &   17 & 8.4 & 8.99 (0.64) & 6.73 (0.20) & 0.531 (0.101) & 4.8 (3.3) & 0.36 (0.30) & 3.0 (1.7) & 1.06 (1.01) & 0.18 (0.15) & 0.67 (0.03) \\ 
 Mk5 &   15 & 8.9 & 8.21 (0.52) & 5.23 (0.62) & 0.018 (0.010) & 4.1 (3.2) & 0.29 (0.25) & 4.0 (2.6) & 0.89 (0.92) & 0.15 (0.14) & 0.30 (0.16) \\ 
 CG10 &    8 & 0.4 & 7.73 (0.90) & 5.45 (0.81) & 0.027 (0.021) & 4.7 (3.3) & 0.44 (0.32) & 3.6 (2.4) & 0.87 (0.86) & 0.14 (0.14) & 0.20 (0.24) \\ 
 IZw18 &   12 & 0.7 & 7.20 (0.64) & 5.85 (0.30) & 0.088 (0.027) & 3.6 (3.0) & 0.33 (0.28) & 5.5 (3.0) & 0.33 (0.02) & 0.23 (0.15) & 0.12 (0.06) \\ 
 Was5 &    9 & 0.4 & 7.87 (0.95) & 5.58 (1.04) & 0.036 (0.034) & 4.5 (3.3) & 0.46 (0.33) & 4.3 (2.8) & 0.72 (0.78) & 0.11 (0.12) & 0.23 (0.28) \\ 
 Mk36 &   14 & 2.5 & 7.02 (1.12) & 5.45 (0.13) & 0.023 (0.004) & 4.0 (3.1) & 0.53 (0.33) & 4.2 (2.7) & 0.33 (0.02) & 0.18 (0.15) & 0.10 (0.08) \\ 
 UM439 &    9 & 0.3 & 7.96 (0.44) & 5.93 (0.73) & 0.099 (0.094) & 5.3 (3.3) & 0.38 (0.30) & 3.7 (2.4) & 0.69 (0.78) & 0.20 (0.16) & 0.15 (0.18) \\ 
 Mk750 &    7 & 0.1 & 7.15 (1.00) & 5.08 (0.93) & 0.011 (0.011) & 4.7 (3.3) & 0.45 (0.32) & 3.9 (2.6) & 0.68 (0.76) & 0.16 (0.15) & 0.46 (0.39) \\ 
 UM461 &   15 & 1.2 & 7.52 (0.56) & 5.22 (0.32) & 0.016 (0.005) & 5.1 (3.3) & 0.45 (0.33) & 4.6 (2.9) & 0.60 (0.59) & 0.17 (0.15) & 0.41 (0.26) \\ 
 UM462 &   16 & 2.2 & 8.60 (0.65) & 5.95 (0.20) & 0.089 (0.018) & 4.2 (3.2) & 0.49 (0.33) & 5.3 (3.0) & 0.75 (0.79) & 0.11 (0.12) & 0.66 (0.03) \\ 
 Mk67 &   10 & 0.7 & 7.88 (0.63) & 5.70 (1.00) & 0.057 (0.061) & 4.8 (3.3) & 0.37 (0.30) & 4.1 (2.7) & 0.72 (0.82) & 0.17 (0.15) & 0.34 (0.30) \\ 
 Mk475 &   15 & 1.4 & 7.47 (0.78) & 4.91 (0.44) & 0.008 (0.003) & 4.1 (3.2) & 0.34 (0.29) & 3.7 (2.4) & 0.92 (0.87) & 0.18 (0.16) & 0.18 (0.13) \\ 
 Mk900 &   15 & 3.6 & 9.18 (0.30) & 5.79 (0.31) & 0.060 (0.017) & 2.5 (2.1) & 0.42 (0.32) & 5.3 (3.0) & 2.09 (0.98) & 0.28 (0.16) & 0.21 (0.06) \\ 
 Mk324 &   10 & 0.3 & 8.69 (0.71) & 6.23 (1.30) & 0.163 (0.192) & 5.2 (3.3) & 0.40 (0.31) & 4.3 (2.7) & 0.88 (0.85) & 0.14 (0.14) & 0.34 (0.35) \\ 
 Mk328 &   10 & 0.4 & 8.87 (0.51) & 6.14 (1.35) & 0.130 (0.160) & 3.9 (3.1) & 0.38 (0.31) & 4.2 (2.7) & 1.17 (1.01) & 0.12 (0.12) & 0.43 (0.40) \\ 
\enddata
\tablecomments{
Bayesian-like uncertainties are given in parentheses following
values. Uncertainties on the logarithm of stellar mass (columns 4 \&
5) are given as fractions of uncertainty divided by value
($\sigma$/value).
}
\end{deluxetable*}

\begin{deluxetable}{cccc}
\centering
\tablecaption{Uncertainties and deviations of derived parameters
\label{err1}}
\tablewidth{0cm}
\tablehead{
\colhead{ \hspace{0.2cm} Param. \hspace{0.2cm} } & 
\colhead{ \hspace{0.2cm} Bayesian \hspace{0.2cm} }  & 
\colhead{ \hspace{0.2cm} Mock \hspace{0.2cm} } & 
\colhead{ \hspace{0.2cm} empirical \hspace{0.2cm} }  \\
\colhead{}      & \colhead{avg. uncert.}    & \colhead{avg. dev.}    & \colhead{comp.} \\
\colhead{(1)}       & \colhead{(2)}              & \colhead{(3)}           & \colhead{(4)}        
}
\startdata
$\langle$SFR$\rangle$$_{10}$ & 0.62 $\pm$ 0.48 & -0.04 $\pm$ 0.25 & \nodata \\  
$\langle$SFR$\rangle$$_{50}$ & 0.58 $\pm$ 0.39 & -0.03 $\pm$ 0.19 & 0.34$^a$ \\  
$\langle$SFR$\rangle$$_{100}$ & 1.00 $\pm$ 0.42 & 0.06 $\pm$ 0.23 & 0.29$^b$ \\  
$\langle$SFR$\rangle$$_{500}$ & 0.55 $\pm$ 0.26 & 0.03 $\pm$ 0.15 & \nodata \\  
$\langle$SFR$\rangle$$_{1000}$ & 0.72 $\pm$ 0.29 & -0.03 $\pm$ 0.19 & \nodata \\  
SFR$_\textrm{all}$ & 0.40 $\pm$ 0.14 & -0.01 $\pm$ 0.07 & \nodata \\  
$M_{\star,o}$ & 0.43 $\pm$ 0.18 & 0.02 $\pm$ 0.10 & 0.14$^c$ \\  
$M_{\star,y}$ & 0.62 $\pm$ 0.48 & -0.04 $\pm$ 0.25 & \nodata \\  
\hline
age$_o$ & 0.54 $\pm$ 0.12 & 0.01 $\pm$ 0.10 & \nodata \\  
$\tau_o$ & 0.66 $\pm$ 0.20 & -0.00 $\pm$ 0.13 & \nodata \\   
age$_y$ & 0.90 $\pm$ 0.21 & -0.01 $\pm$ 0.17 & \nodata \\  
$\tau_y$ & 0.76 $\pm$ 0.13 & -0.02 $\pm$ 0.12 & \nodata \\  
$f_\textrm{burst}$ & 0.88 $\pm$ 0.23 & -0.05 $\pm$ 0.15 & \nodata \\  
$A_\textrm{FUV}$ & 0.35 $\pm$ 0.38 & 0.01 $\pm$ 0.08 & \nodata \\  
$A_\textrm{V}$ & 0.40 $\pm$ 0.36 & 0.00 $\pm$ 0.10 & \nodata \\  
$\alpha$ & 0.15 $\pm$ 0.13 & 0.02 $\pm$ 0.05 & \nodata \\  
$\delta$ & 0.45 $\pm$ 0.21 & -0.03 $\pm$ 0.12 & \nodata \\  
\enddata
\tablecomments{
Column 2: Average (and rms) of uncertainties from
   Bayesian-like analysis of PDFs 
   (in dex).
Column 3: Average (and rms) of deviations from mock analysis (in dex).
Column 4: Scatter in comparison to observables (in dex): 
$^\textrm{a}$ from \Ha observations, 
$^\textrm{b}$ from UV observations, 
$^\textrm{c}$ for total stellar mass from $3.6\mu$m observations,
Values are included in these averages for the 257 real
fits and 162 mock fits with measured
   fluxes in all filters and  $\chi_\textrm{red}^2 < 5$.
}
\end{deluxetable}

In addition to these internal consistency checks on the SED grids and
fitting methods, we can also use comparisons with independent
determinations of similar parameters to verify our results. Most
importantly, we can compare our SED-derived SFRs with those
determined from \Ha observations. We can also compare the SED-derived
SFRs with estimates from the UV flux and compare the SED-derived
stellar masses with estimates from $3.6\mu$m flux and K$_\textrm{s}$
flux. These are less independent
comparisons than in \Ha as these fluxes are already
included in the SED fits. Still, these comparisons are an
important consistency check of our fits, and also help provide a
pathway toward comparing our SED-derived parameters with even broader
samples from the literature.

\Ha fluxes have been used to estimate SFRs for
the LVL galaxies in \citet{lee09} as part of the 11HUGs project. We use
their extinction-corrected \Ha SFRs, which have been
determined using the standard \citet{kennicutt98} relationship with a
dust correction as
described in \citet{lee09}, and follow a similar procedure for the BCD
\Ha fluxes. Figure~\ref{sfr} shows the comparison between our SED SFRs
and the SFRs from \citet{lee09} for the galaxies in common, and the
same comparison using UV photometry. The \Ha emission is most
sensitive to star formation occurring over 
the past $\sim$$10$~Myrs while the shortest meaningful time
scale we can measure SFR in the SED fits is $\sim$$50$~Myrs. Despite
these timescale differences, the
agreement is still quite good. The UV estimates of SFR are in better
agreement with the 100~Myr-averaged SFH since the timescales are more
closely matched, but the UV fluxes are themselves included in the
SED-fitting.

The relationship between SED and \Ha SFRs for the LVL galaxies
shows generally good agreement, with a scatter of $\sim$0.3~dex. The
\Ha fluxes are not 
included in the SED fits, as it remains a challenge to incorporate
them at low redshifts, but is often required for galaxies at 
higher redshifts \citep{ono10, stark13, debarros14}. In our case, the
\Ha fluxes give an independent estimate of the recent star formation
in our sample, and allow us to assess the reliability of our SED
fits. 
However, the SED fits rely primarily on UV fluxes to determine SFRs,
while the \Ha SFRs are measuring somewhat different star formation
activity (see Section~\ref{extremeMZ} for further discussion
on the systematic offset between SFRs from \Ha fluxes and SED
  fits for the BCDs).

Consistent with many previous studies, \citet{McGaughSchombert14} find
that NIR luminosities of galaxies are a good tracer of
their stellar mass, almost independently of their color (e.g.,
evolutionary state). They give mass-to-light ratios of M$/$L$=$$0.6$
$M_\odot/L_\odot$ in the K$_s$ filter, and M$/$L$=$$0.47$$M_\odot/L_\odot$
in the $3.6\mu$m band of Spitzer-IRAC (using Vega magnitudes). We use
these mass-to-light 
ratios and our observed photometry to determine stellar masses for
our galaxies, shown in Figure~\ref{M}. For BCDs without Spitzer
observations, we generate $3.6\mu$m stellar masses from \textit{w1}
band photometry from the
ALLWISE\footnote{\url{http://wise2.ipac.caltech.edu/docs/release/allwise/}}
 catalog from the  Wide-field
Infrared Survey Explorer \citep{wright10}.

\begin{figure*}[htb]
\centering
\includegraphics[width=0.47\textwidth]{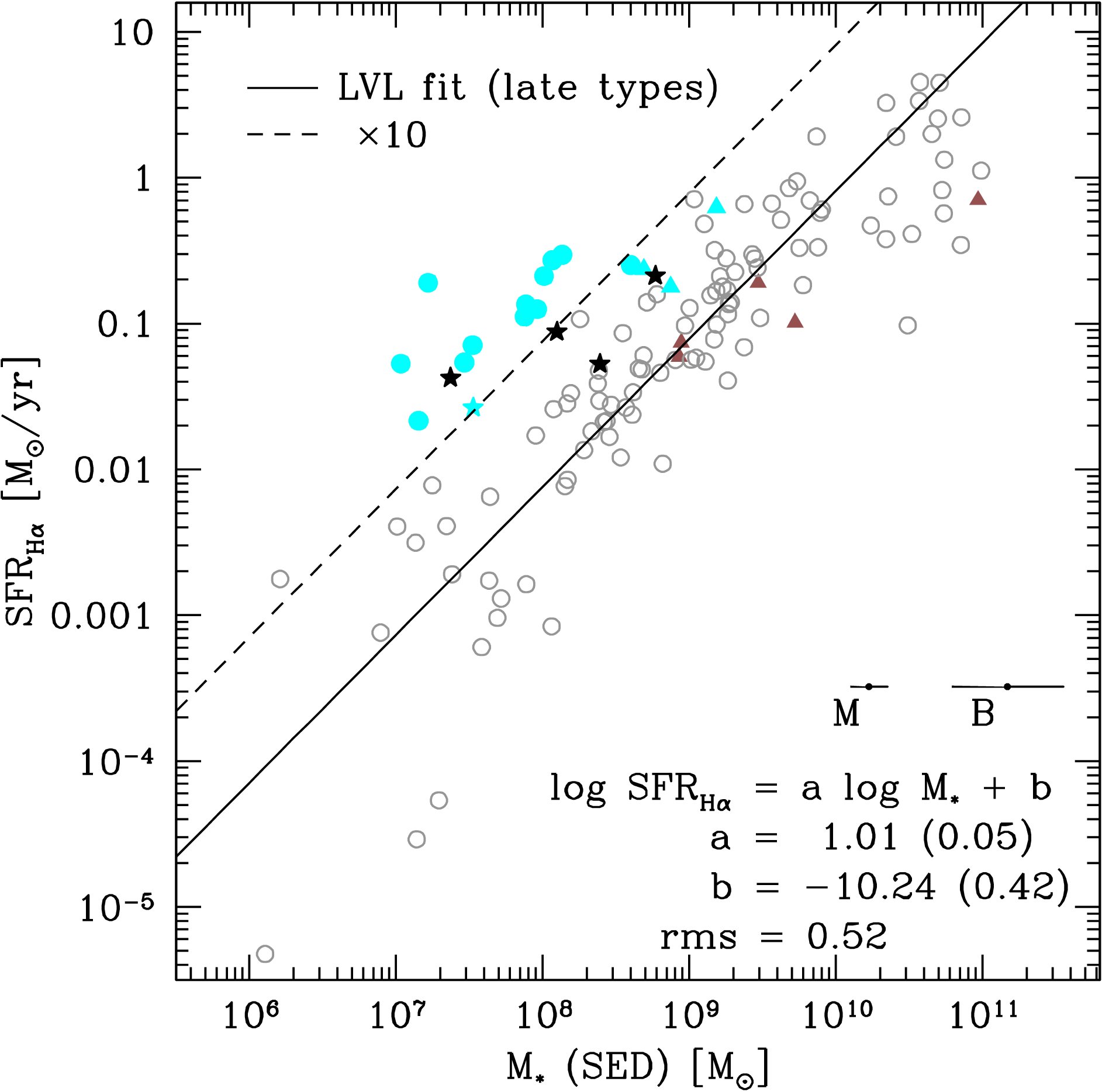}
\hspace{0.2cm}
\includegraphics[width=0.47\textwidth]{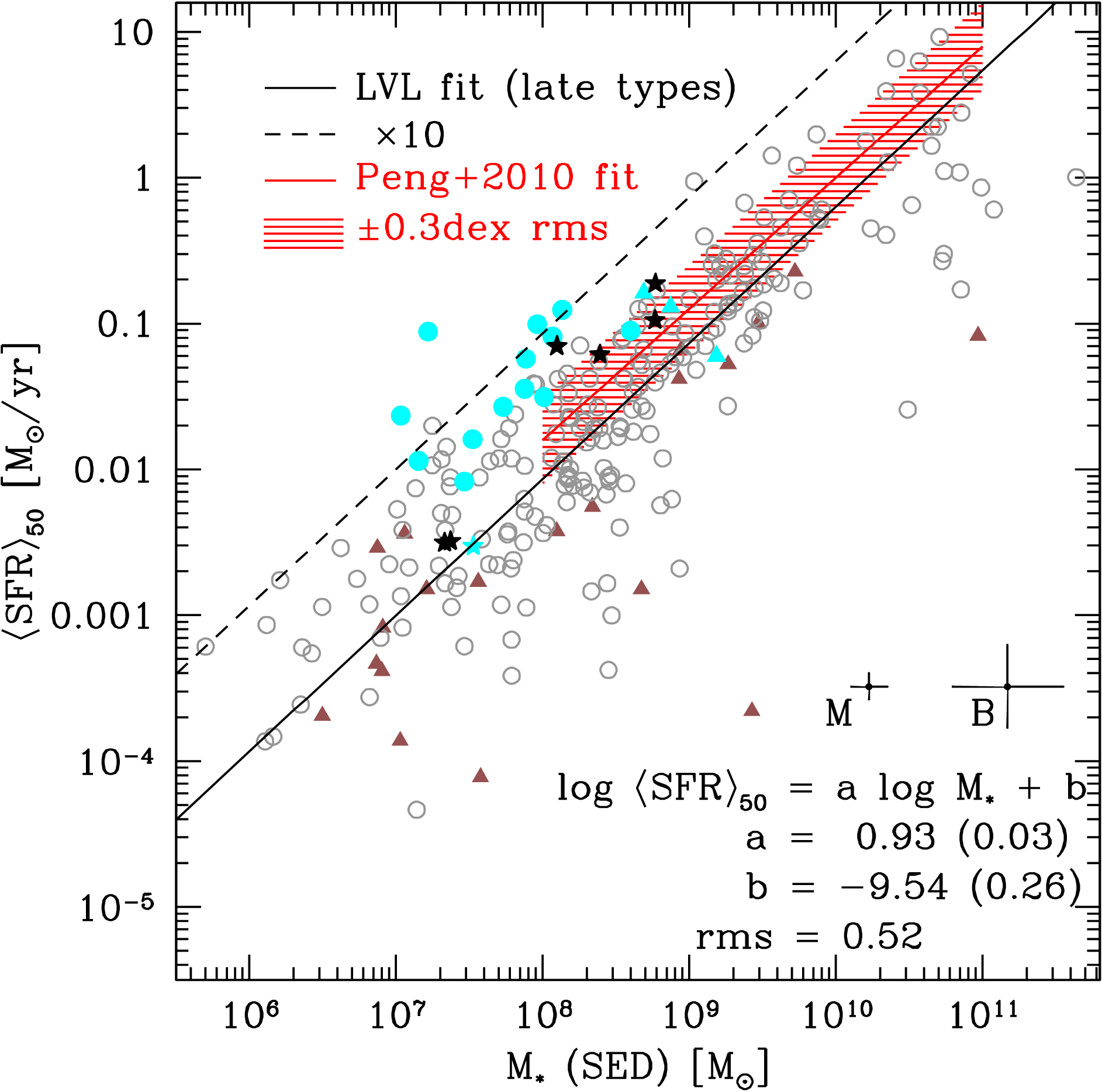}
\caption[SFR vs $M_\textrm{stellar}$]{SFR versus stellar
  mass for the complete LVL and BCD samples (color-coding is the same
  as Figure~\ref{sfr}.
The left panel shows the SFR determined from \Ha observations (for the
  galaxies with \Ha measurements) while the
  right panel shows the $50$-Myr average SFRs determined from the SED
  fits.
Typical error bars are shown from mock (``M'') and Bayesian (``B'')
   analysis. 
The red shaded area corresponds to the relationship found by
   \citet{peng10} for $z$$\sim$$0.1$ SDSS galaxies.
Both panels show the best-fit line from the late-type LVL galaxies,
  and its slope ($a$), intercept ($b$), and their uncertainties. The
  BCDs (from our sample and the LVL sample) have higher SFRs than
  typical galaxies of their mass, and are especially extreme in
  SFR$_\textrm{\Ha}$.
\label{MSFR}}
\end{figure*}

Figure~\ref{M} shows the generally good agreement between SED masses
and luminosity-based estimates. This is unsurprising as the SED fits
already use the same fluxes that are used to generate the
masses via mass-to-light ratios. Still, these comparisons demonstrate
that our SED-fitting methods produce stellar masses which are
consistent with those found using simpler methods. Further, the 
small offset and lack of mass-trends in the relationships
indicate that there are no 
significant mass-dependent systematic effects.

\subsection{Total uncertainty budget}

Putting all of these reliability and uncertainty estimates together,
we now arrive at a complete error budget for our SED fit
results, shown in Table~\ref{err1}. This includes the Bayesian-like
uncertainties from the SED-fits, the mock analysis of reliability, and
the external comparison with observations. In general, the
uncertainties are small enough that it is possible to fit the SEDs of
BCDs and LVL galaxies and 
reliably determine some of their physical parameters. In particular,
we find that our estimates of stellar mass have a typical Bayesian
uncertainty of $\sim$$0.5$~dex, but agree with empirical comparisons
within $0.14$~dex. Similarly our estimates of SFR on various time
scales have formal Bayesian uncertainties between $0.40-1.00$~dex,
but show significantly better agreement with empirical comparisons of
$0.25-0.30$~dex, and mock deviations that are even smaller.
These SED-derived parameters are not ``high precision'' 
measurements, but are constrained well enough to allow for meaningful
comparisons between the BCD and LVL samples. These uncertainty
estimates are included on all subsequent plots, and demonstrate the
reliability and accuracy of the parameters.


\section{Results from SED fits}\label{sed_results}

%
%
%
Having generated best-fit parameters for the full sample of BCDs 
(shown in Table~\ref{best1}) and LVL galaxies, we can now compare
these populations. To put the BCDs
(from our sample and the LVL sample) 
into a broader context, we consider the non-BCD late-type
galaxies in LVL as ``normal'' galaxies. We are
interested in testing whether 
BCDs are best described as a phase of dI evolution, or whether BCDs
represent a unique type of dwarf galaxy. While some of the SED-derived
parameters are difficult to
accurately and reliably determine, our fitting method is uniformly
applied to the observations of the BCDs and LVL galaxies, so
we can study the 
statistical trends between the two populations in a differential
sense. In the following section 
we discuss the results of our SED fitting and the implications they
have on the evolutionary context of BCDs.

\begin{figure*}[htb]
\centering
\includegraphics[width=0.47\textwidth]{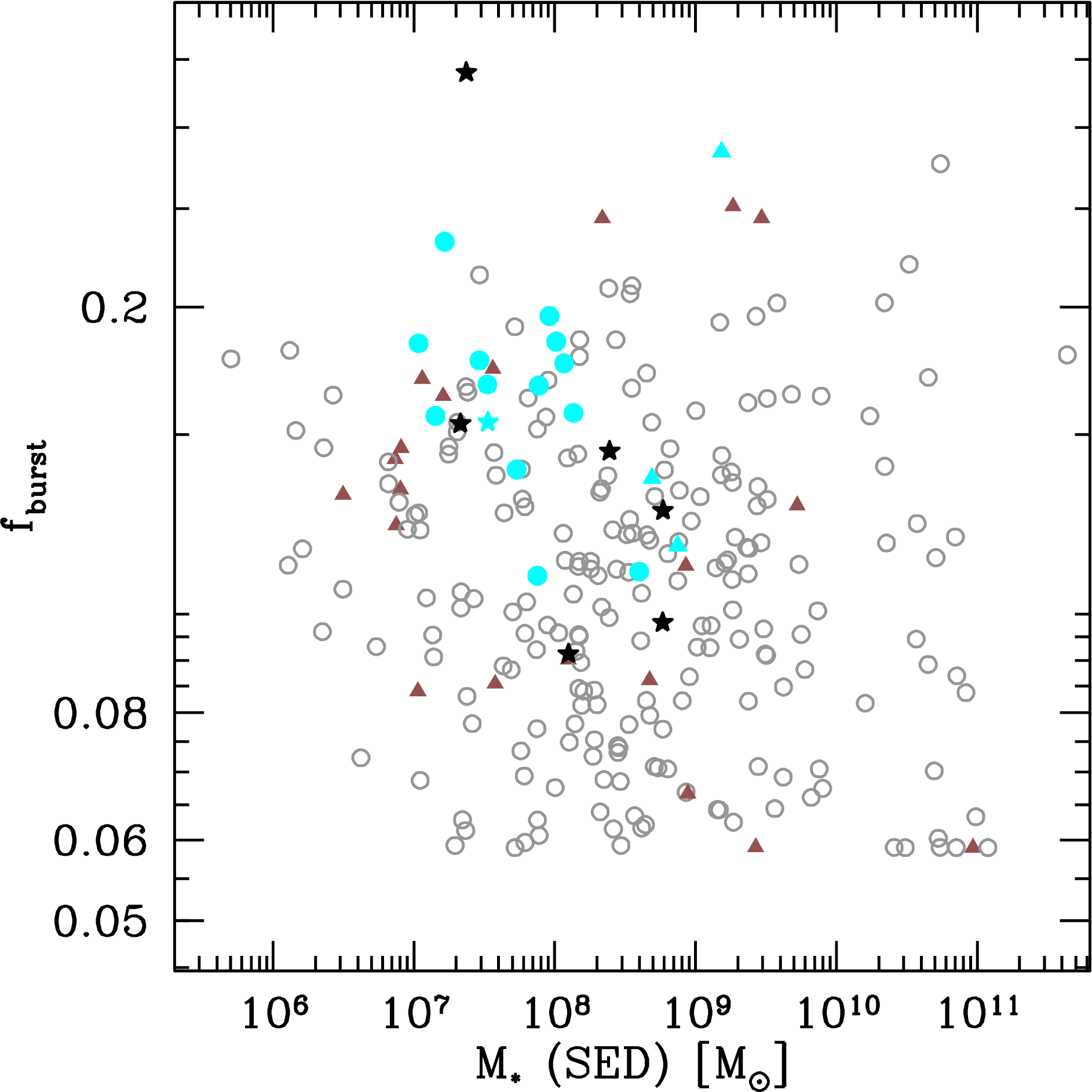}  
\hspace{0.2cm}
\includegraphics[width=0.47\textwidth]{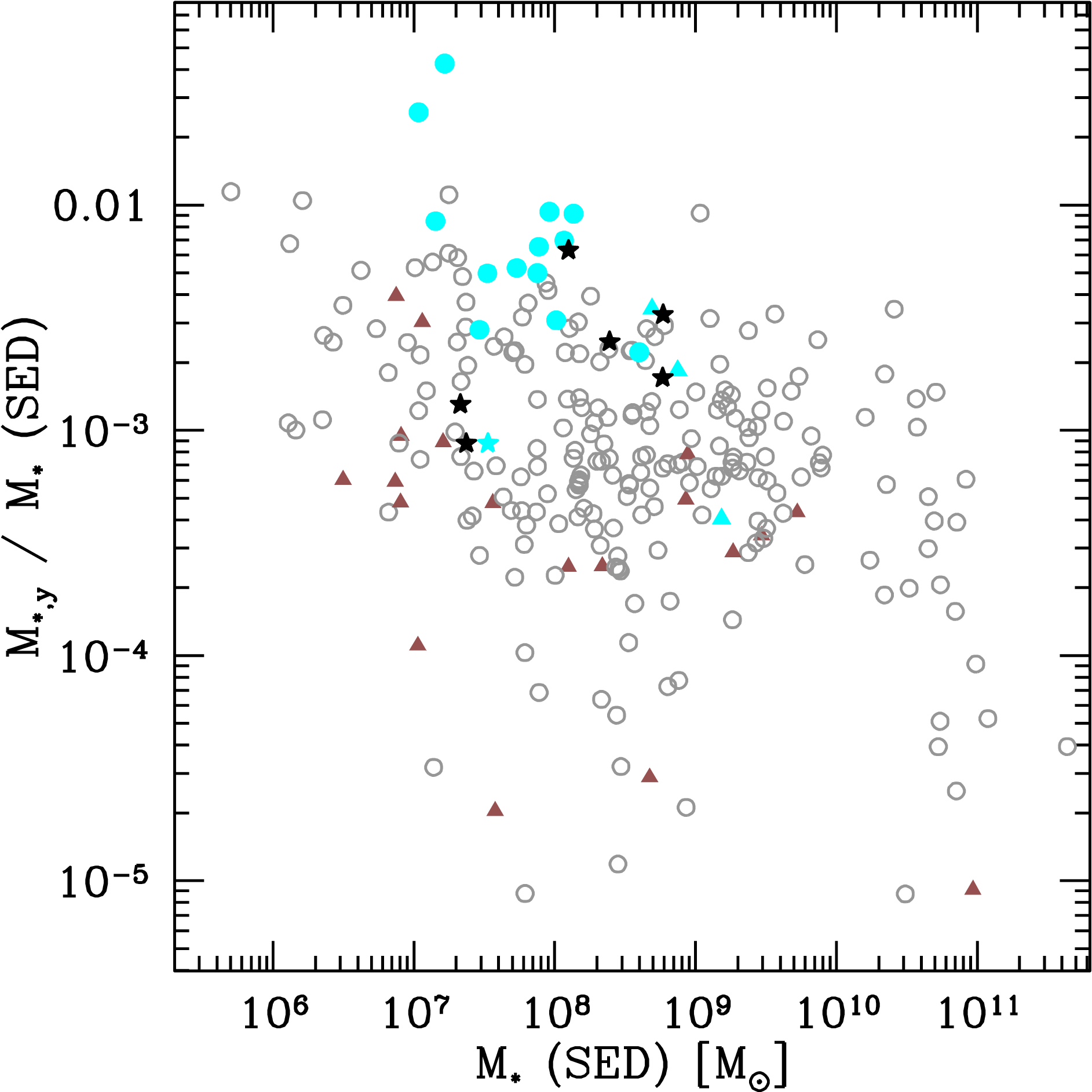} \\
\caption[Comparison between burst estimates and
  $M_\textrm{stellar}$]{Left panel shows the relationship between
  $f_\textrm{burst}$ 
  (initial mass ratio of young population) and total stellar
  mass. Right panel shows the relationship between the present day
  young stellar population mass ratio as a function of total stellar mass.
Color-coding is as in Figure~\ref{sfr}.
When compared with late-type LVL galaxies of similar total stellar
  mass, BCDs have more extreme young stellar population mass ratios
  (right panel) than burst fractions (left panel).
\label{bfracM}}
\end{figure*}

\subsection{Stellar Mass vs SFR relation}
\label{MSFRsec}

Figure~\ref{MSFR} shows two versions of the
M$_\textrm{stellar}-$SFR~diagram, where the 
the late-type LVL galaxies populate the so-called ``main sequence'' of
star-forming galaxies. We show
measurements of the SFR from the SED fit averaged over 50~Myr and from
\Ha observations. As found by previous
studies \citep[e.g., the shaded area from][]{peng10}, there is a
relatively tight relationship between SFR and stellar mass. The BCDs
lie at (and 
above) the upper extreme of this  
relationship, in the sense that they have larger SFRs than most other
galaxies of their stellar mass.

The relatively tight M$-$SFR relationship is typically interpreted as
evidence of a universal
mode of star formation in 
star-forming galaxies, where the
global amount of star formation in a galaxy is regulated (in some way)
by its stellar mass.
%
%
 This strong correlation between SFR and
M$_\textrm{stellar}$ is seen across five orders of magnitude of
mass. As discussed more extensively in \citet{cook14b}, there is good
agreement in the overlap between the relationship from the late-type
LVL galaxies and that of \citet{peng10}. The early-type LVL galaxies
are typically offset below the best-fit line, as they have lower SF
activity than comparable late-type galaxies.
The LVL sample extends to
significantly lower stellar masses than the SDSS sample, and confirms
that the star-forming main sequence relationship continues at least
down to M$_\textrm{stellar}$~$\sim$$10^6M_\odot$, although it may
become broader at the lower masses.

As seen in Figure~\ref{MSFR}, regardless of whether SFR is measured
via the 50-Myr average from SED fits or from \Ha fluxes, the M$-$SFR
relationship for BCDs is offset from that of normal
galaxies. When considering SFR$_\textrm{\Hans}$ measurements, the BCDs are
more than an order of magnitude offset from the typical relationship,
in the sense that BCDs have $\sim$$10$ times more star formation
per unit stellar mass. When using $\langle$SFR$\rangle_{50}$, the offset between the
BCDs and normal galaxies is smaller, but is still nearly an order of
magnitude. 
Regardless of
the implicit assumptions in different SFR indicators, BCDs
show SFRs elevated above the
star-forming ``main sequence'', indicating that BCDs are
experiencing truly exceptional amounts of star formation. As their \Ha
SFRs are elevated beyond the SED SFRs (based on UV fluxes), the
current SF in BCDs may also be especially recent.

\subsection{Burst strength}
\label{bstrength}

As part of its SED-fitting process, CIGALE determines the ``burst
fraction'' ($f_\textrm{burst}$) for each galaxy, defined as the
initial mass fraction of the young stellar 
population. More explicitly, this is the mass fraction of the young
stellar population compared to the total stellar mass, \emph{which are
  both taken at the
  time of the young population's birth.} The present-day mass ratio of
the young stellar population will always be smaller than the initial
$f_\textrm{burst}$ as the young high mass stars are perishing at a
faster rate than the older stellar populations.

Figure~\ref{bfracM} shows both the initial $f_\textrm{burst}$ and the
present-day young stellar
population mass ratio as functions of total stellar mass. 
There is a substantial overlap between the late-type LVL
galaxies and BCDs in terms of their initial $f_\textrm{burst}$ values (left
panel), which may mean that BCDs and non-BCDs may have experienced
similarly strong starbursts in their past.
However, BCDs with 
lower stellar masses typically have higher initial $f_\textrm{burst}$
values, which is consistent with a single star formation event having
a more
significant effect on a lower mass galaxy. One of the BCD/E galaxies,
Mk~900, has $f_\textrm{burst}=0.28$ and is an exception to this trend.
Its young
stellar population is the oldest among the BCD sample fits
(age$_y$=2~Gyr), so its burst strength is also the most extreme. Note
that its young population mass ratio ($M_{\star,y}/M_\star$) is
$\sim$$10^{-3}$, which is typical for other galaxies of its mass.

While there is overlap between BCDs and late-type LVL galaxies, the
BCD population is offset higher than normal galaxies, suggesting that
the star forming events (in terms of the mass involved) may be more
substantial in BCDs.

When considering the young stellar population mass ratio (shown in the
right panel of Figure~\ref{bfracM}), the BCDs more clearly populate
the extreme of the parameter space. While BCDs and non-BCDs
galaxies may have had more similar values of initial burst
strengths (left panel), the mass fraction of today's young stellar
population is much greater in BCDs than in LVL galaxies (right
panel). Much of this difference is due to the age of their young
stellar populations. BCDs have a significant young star-forming
population, while the stellar populations in non-BCDs have had more
time to age and fade. Again there is a
trend for BCDs with lower stellar masses to have larger young
population mass fractions. 

Combined, the two panels of
Figure~\ref{bfracM} suggest that the young stellar populations of
BCDs are different from normal late-type galaxies today. Their
SED-derived burst strength does not separate them as clearly, but
their young population is usually more significant. Furthermore, the
exponential scale times associated with the young and old stellar
populations can vary, which can add to the differences between these
comparisons.

\begin{figure}[tb]
\hspace{-0.7cm}
\includegraphics[width=0.47\textwidth]{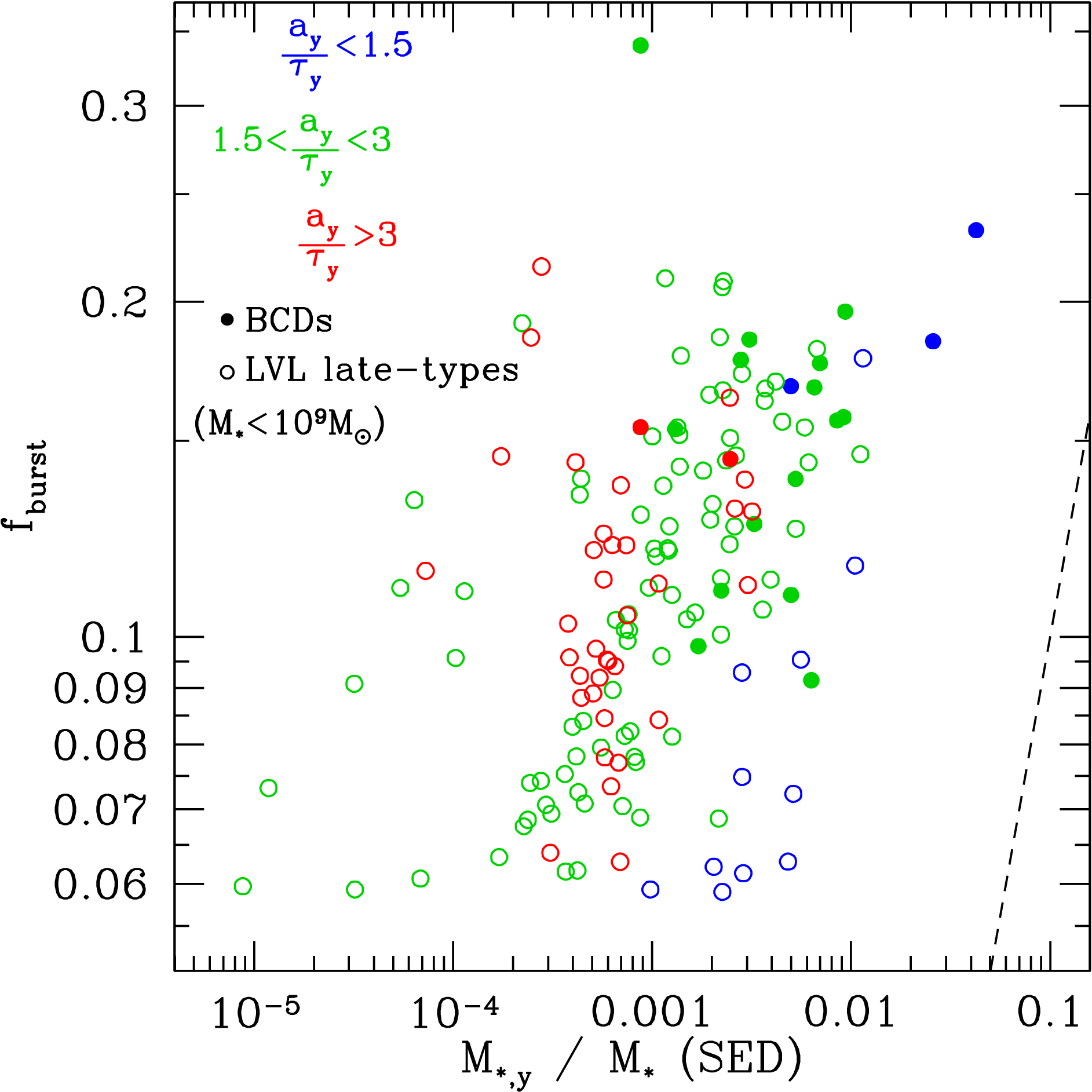} 
\caption[Comparison of $f_\textrm{burst}$ with young mass
  fraction]{Comparison between best-fit $f_\textrm{burst}$ values and 
  estimates of the current mass ratio of the young stellar
  population.
Each galaxy is plotted with a shape and color that correspond to the
  age of its young burst population, in units of its scale time,
  $\tau$. Galaxies with bursts of age/$\tau$$<$1.5 are shown in blue,
  middle-aged bursts are shown in green, and older bursts are shown in
  red.
Solid dots show BCDs (from our sample and LVL) and open circles show
  late-type LVL galaxies, selected to be in the same stellar mass
  range as the BCDs.
The dashed diagonal line shows the unity line, which indicates a very
   recent burst.
The galaxy with the highest $f_\textrm{burst}$ value is UGCA~281, an
  LVL galaxy classified as a BCD and discussed further in
  Section~\ref{bstrength}. The galaxy with the highest M$_{*,y}$/M$_*$
  is I~Zw~18.
In general, BCDs are characterized by having both strong burst strength and a
  significant young population at present day, although there is
  substantial scatter.
\label{bfrac}}
\end{figure}

Further insight can be gained by combining these two estimates of
burst strength into Figure~\ref{bfrac}, which shows $f_\textrm{burst}$
as a function of the young stellar population mass ratio. The points
on Figure~\ref{bfrac} are color-coded by their young population 
burst age (age$_y$) divided by its exponential scale time
($\tau_y$). Blue points show galaxies with young populations younger
than $1.5\tau$, which are very recent bursts. Green points show
middle-aged bursts, and red points show 
more evolved bursts. We plot all of the BCDs as dots (from our sample
and the LVL sample) and the LVL late-type galaxies as circles. We only
show galaxies from LVL in the stellar mass range
(M$_\textrm{*}<10^9$\msun), to match the BCD sample.

If the young population bursts were extremely recent, the points would
lie very close to the equality line (shown as a dashed 
diagonal line, where
f$_\textrm{burst}$=M$_\textrm{*,y}$/M$_\textrm{*}$). As the young
population evolves, the points move horizontally 
in this plot as the young stellar 
population mass ratio decreases (while its initial burst fraction
always remains the same). This is seen on the plot as the points
closest to the 1:1 line have younger bursts (i.e. are blue points),
although there is significant scatter beyond this simple
trend due to different age$_y$/$\tau_y$ parameters of the SFHs. Compared
with the late-type LVL galaxies, on average the BCDs 
show higher values of both $f_\textrm{burst}$ and young
population mass ratios, as expected. 

With this relationship in mind, it appears that the star formation in
BCDs is both substantial and typically more recent (than LVL). Note
that some of the late-type galaxies in LVL have 
similar young populations to BCDs. The requirements of recent and
substantial young stellar population are necessary, but not
sufficient, criteria for a galaxy to be a BCD.

It is interesting to note that Figure~\ref{bfrac} also shows a
population of LVL galaxies with moderate initial
$f_\textrm{burst}$ (e.g., $>0.1$) but with low present-day young
population mass fraction (e.g., $< 0.5\%$). These characteristics are
similar to what we would expect 
from a population of post-burst BCDs. When the young populations in
those galaxies with high $f_\textrm{burst}$ first started forming stars,
by definition their young population mass fraction would have been
equivalent to $f_\textrm{burst}$. While we do not have any objects in
our sample with young population mass fractions as extreme as $10\%$,
galaxies in this state would unambiguously be considered starbursting
systems (and possibly BCDs).

\begin{figure*}[htb]\centering
\includegraphics[width=0.45\textwidth]{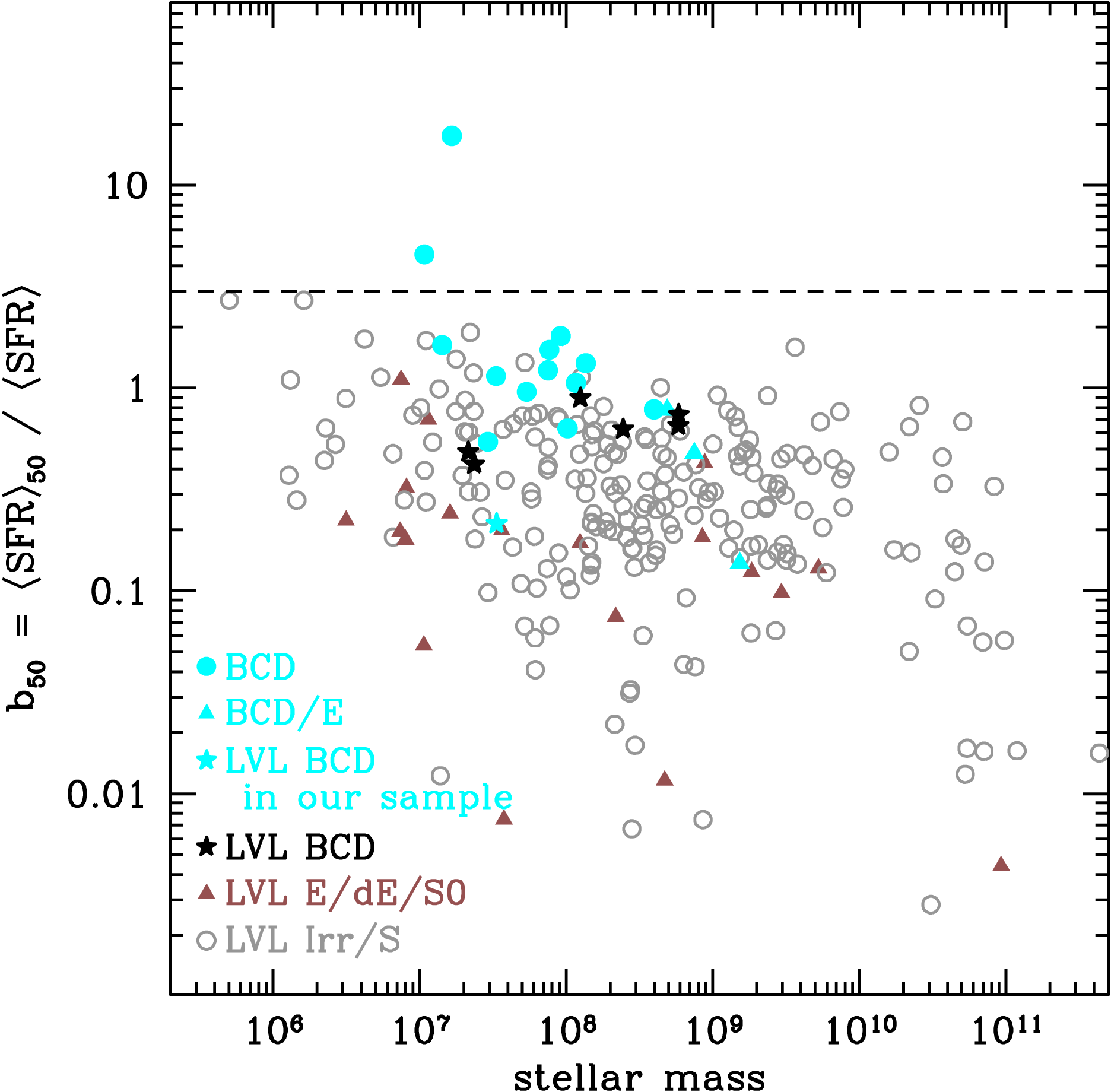} 
\includegraphics[width=0.45\textwidth]{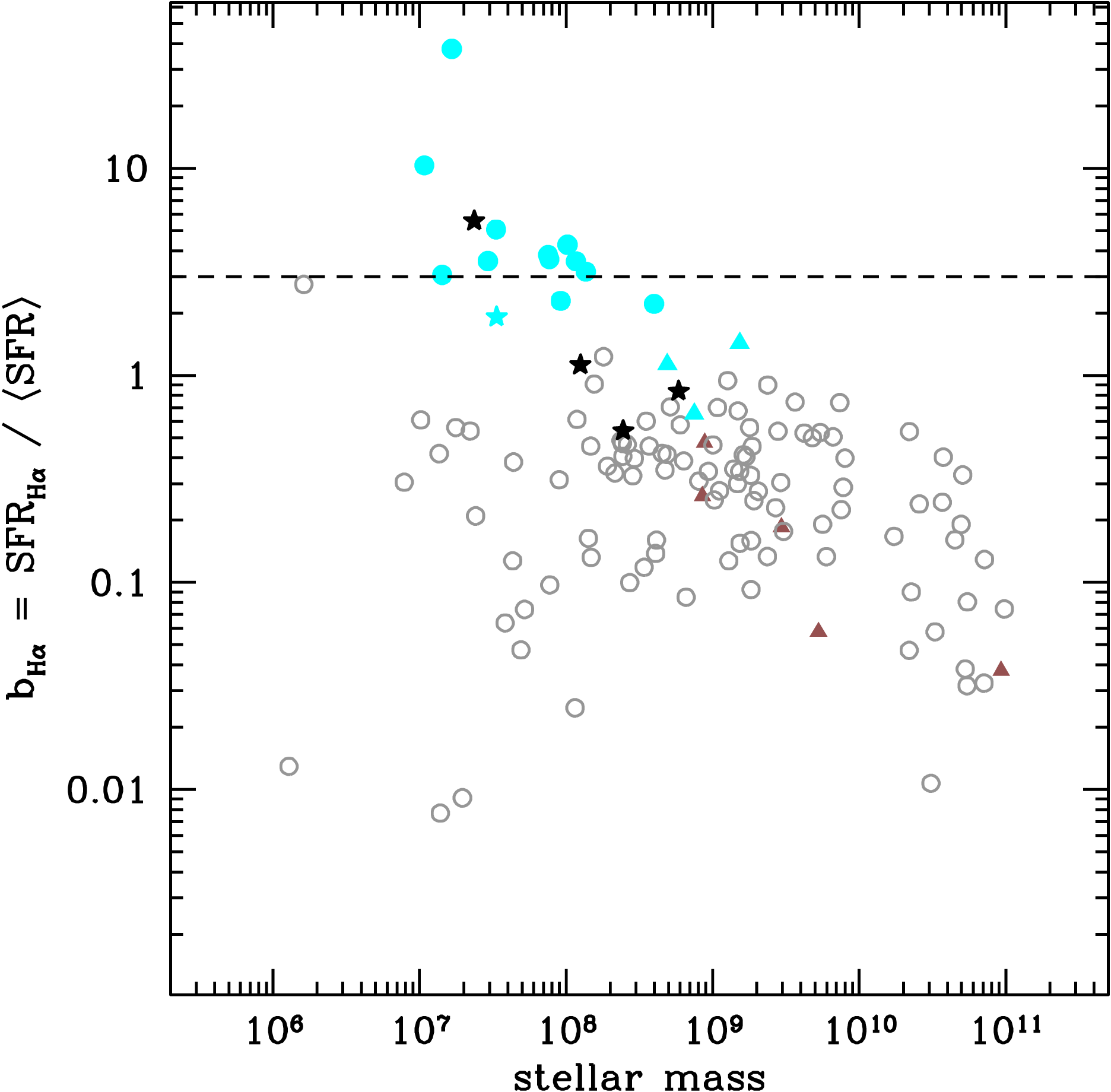}
\caption[Birthrate parameter compared with stellar mass]{
  Left panel
  shows the 50-Myr averaged birthrate ($b_{50}$), 
  and right panel shows the birthrate based on SFR$_\textrm{\Hans}$
  ($b_\textrm{\Hans}$), with color scheme as earlier.
Very few galaxies meet the $b \ge 3$ threshold for starbursts when
  using SFR$_{50}$ from the SED fits, but the
  \Hans-determined SFR shows many of the BCDs above the
  starburst threshold. The
  BCDs have higher birthrate parameters than the late-type LVL
  galaxies, but some LVL galaxies also have large values of
  $b$.
\label{bSFR}}
\end{figure*}

The young population burst ages of the BCDs (shown in
Table~\ref{best1}) 
are somewhat larger than might be expected. Some of the most
extreme BCDs have young population ages $< 500$~Myr (e.g., I~Zw~18 and
Mk~36), but 
many have young population ages between 500 and 1000~Myr, and some even
$>1000$~Myr. However, based on the strong \Ha emission from the BCDs,
it is unlikely that their young populations can be this
old. Instead, this over-estimate of burst age demonstrates a
limitation of our SED fits. The broadband fluxes used in the SED fit
are most sensitive to star formation on 100-1000~Myr timescales, but
only crudely and with low resolution. Even if the BCDs have had recent
(50-100~Myr) star formation events, the SED fits are unable to resolve
SFH changes on such short timescales. In this way, the SED fits are
almost certainly
systematically under-estimating the current star formation rates of
galaxies which have recently started forming stars, which is
consistent with the behavior seen in Figure~\ref{MSFR}.

As an alternative parametrization of the significance of the current
star formation episode, we also calculate the birthrate parameter of
\citet{kennicutt83}, defined as:
\begin{equation}
b = \frac{\textrm{SFR}}{\langle\textrm{SFR}\rangle}
\end{equation}
\noindent
where $b$ is the birthrate parameter, SFR is an estimate of the
current SFR, and 
$\langle$SFR$\rangle$ is the total lifetime average SFR. 
This parameter is related to the specific star formation rate and also
to the young stellar population mass fraction and $f_\textrm{burst}$,
but now parametrizes the starburst in terms of its star formation rate
instead of the mass of the young stellar population .

We make two versions 
of $b$, one calculated using the most recent 50-Myr averaged SFR
($b_{50}$) as the numerator and one with the SFR determined from \Ha
observations ($b_\textrm{\Hans}$) as the numerator. In both cases, the
denominator is the lifetime average SFR from the SED fits (closely
related to the total stellar mass). 
 These birthrate parameters 
are plotted as a function of stellar mass in Figure~\ref{bSFR}. 
As discussed previously, the \Ha SFR is more sensitive to recent
intense star formation, while the SED SFR probes longer time scales.



Note
that these birthrate parameters are not measuring the stellar mass
involved in the star formation event, but represent the significance
of the current star formation in comparison with a galaxy's total
lifetime of star formation. For a particular galaxy, the birthrate
parameter describes the significance of its current star formation in
its long-term evolution. A galaxy with a birthrate parameter of $b \ge
3$ is typically required to be considered a starburst
\citep[e.g.,][and references therein]{bergvall15}. In the
  smallest dwarf galaxies, the effects of star formation can be more
  significant (i.e., the effects of feedback on galaxy-wide scales)
  than the same absolute amount of star 
  formation in larger galaxies. The birthrate parameter gives a good
  indication of the significance of the current SFR. Starbursting
  galaxies with $b > 3$ typically are undergoing a truly
  transformative event, distinct from the low-level fluctuations in
  SFR which are typical in dwarf galaxies.

In Figure~\ref{bSFR}, the $b$ values for the BCDs show the
correlation with stellar mass in the sense that lower stellar masses
have larger $b$ values, in the same way as $f_\textrm{burst}$ and the
young population mass fraction increased at lower stellar masses. This
behavior is expected if the starbursts are similar in size, but can
have a greater
impact on a lower mass galaxy than on a higher mass galaxy. When
comparing the $b$ values for the BCD and LVL samples, the BCDs have
values which are at or beyond the maximum values from the LVL
non-BCD samples. 
In particular, one LVL galaxy also populates this extreme area:
UGCA~281, which has long been classified as a BCD
\citep{tm81, lelli14}. On the other hand, some members of
our BCD sample (notably the three most massive galaxies: Mk~324,
Mk~328, and Mk~900) 
populate the more normal areas of parameter space.
The early-type galaxies from
the LVL sample (red triangles) and the BCD/Es (cyan triangles) have
some of the lowest values. These are also at somewhat higher masses,
so a star formation event may have a smaller impact on their
evolution.

Whether measured by the 50-Myr average from
SED fits or by \Ha fluxes, the birthrate parameters for BCDs are
extreme, and show that the current star formation in BCDs is
proceeding at a level which is not typical for normal star-forming
galaxies.


\section{Discussion}\label{sed_discussion}

While our SED fits can only generate coarse estimates of key physical
parameters (e.g., M$_\star$, SFR) for the complete LVL sample
and our BCDs, they 
still provide a systematic and meaningful way to compare the BCDs with
normal LVL galaxies. 
Our SED fits have shown that BCDs are forming stars at exceptionally
intense rates and may populate an elevated relation parallel to the
``star formation main sequence.''
However, the SED fits cannot provide the
detailed star formation histories which would be needed to clearly
identify both past and present intense (but unsustained) levels of
star formation in order to separate BCDs (both past and present) from
normal dIs. Much of the detailed 
information about a galaxy's SFH is not accessible through SED
fits alone, but can only be probed via studies of resolved stellar
populations \citep[e.g.,][]{mcquinn15c}. However, our SED fits can
still be used to discuss and constrain the possible
evolutionary connections between dwarf galaxies.


\subsection{SFR and stellar mass indicators at the extremes}
\label{extremeMZ}

Most of the empirical SFR and stellar mass indicators
\citep[e.g.,][]{kennicutt98, McGaughSchombert14} have been 
calibrated using galaxies with large stellar masses
(M$_\star$$\approx$M$_\textrm{MWG}$), with modest 
abundances ($Z$$\approx$$Z_\odot$), and star formation rates
(SFR$\approx$$1$$M_\odot/$yr). More recently, some groups have been
able to test whether these standard relations apply equally well to
galaxies with lower masses, lower abundances, and possibly higher
SFRs.

\begin{figure}[tb]
\centering
\includegraphics[width=0.90\columnwidth]{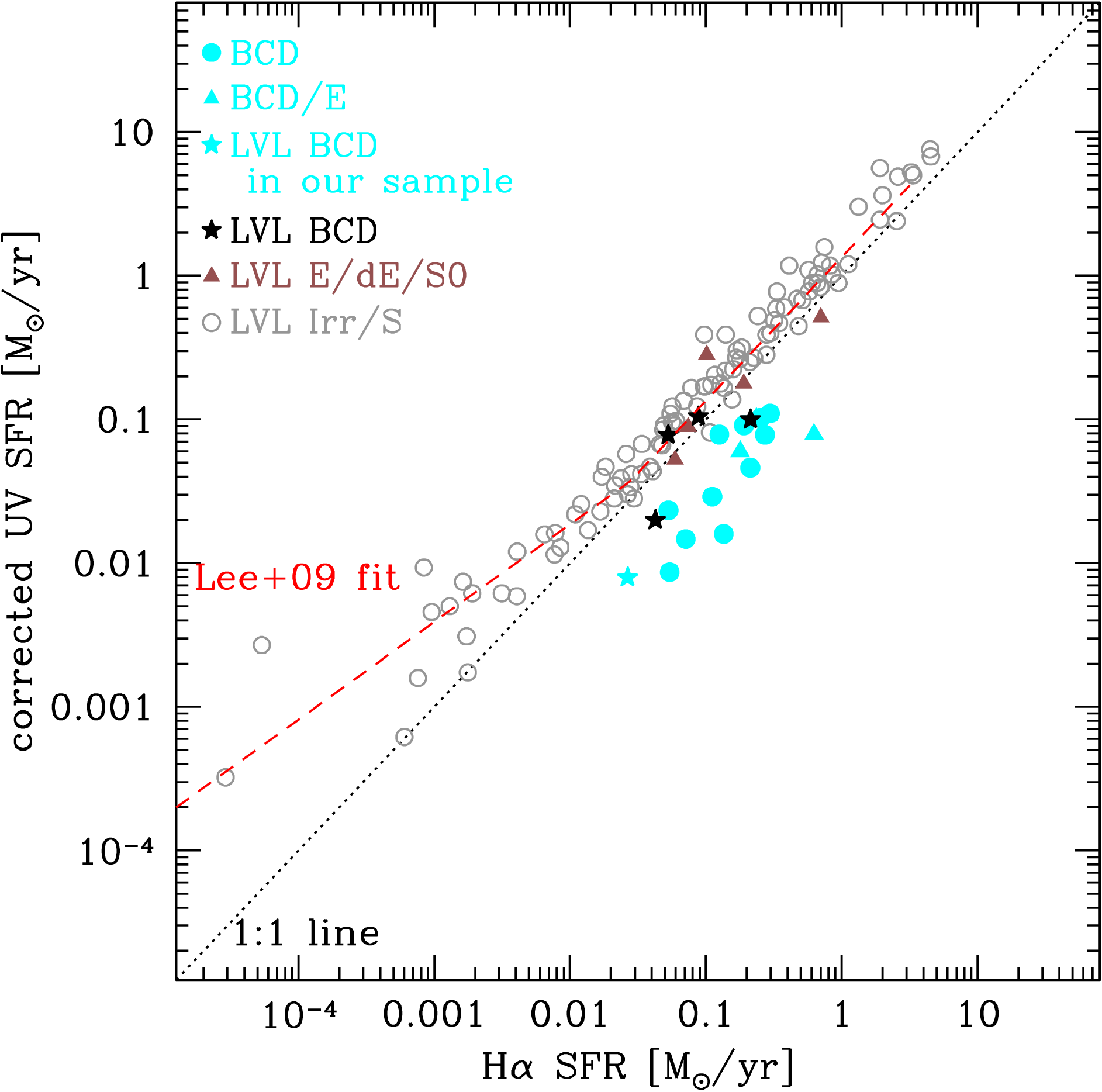} 
\caption[Comparison between SFRs from \Ha and UV fluxes]{This figure
  shows the correlation between observed SFRs from 
  \Ha fluxes and UV fluxes (corrected for internal extinction).
LVL galaxies are shown in black, and BCDs in blue. 
 In black is the equality line. 
The offset between BCDs and LVL galaxies is likely due to assumptions
  in the empirical conversion factors about SFH, which the BCDs do not
  obey.
The red line shows the best fit relation from \citet{lee09}.
\label{hauv}}
\end{figure}

\citet{lee09} found that for
galaxies with low SFR, \Ha 
measurements give systematically
  lower SFRs than are 
 measured in the UV. Similar findings have been reported by
other groups as well \citep{fumagalli11, eldridge12, weisz12}. We can
also compare the SFRs for the galaxies in our sample, as shown
in Figure~\ref{hauv}, and plot SFR from \Ha against SFR from UV. The
best fit relation from \citet{lee09} is also shown, and matches the
same general trend as the LVL data.

The BCDs appear offset from the main population
in Figure~\ref{hauv}, 
which is similar to the offsets previously seen in the SFR
comparisons (e.g., Figure~\ref{sfr}). As mentioned in
Section~\ref{external}, the BCDs have systematically larger
SFR$_\textrm{\Hans}$ than $\langle$SFR$\rangle$$_{50}$, while their
SFR$_\textrm{UV}$ and 
$\langle$SFR$\rangle$$_{100}$ estimates are in better agreement. A
similar offset in 
the BCD SFR$_\textrm{\Hans}$ estimates is seen here in Figure~\ref{hauv}.

A possible explanation for this offset lies in the calibration of
these SFR indicators. The
empirical SFR estimates used to generate SFR$_\textrm{UV}$ and
SFR$_\textrm{\Hans}$ both assume a constant SFR. In the FUV, the flux
comes from young 
stars with lifetimes $\sim$$10^8$~years, and
correspondingly measures the SFR on those time scales. The \Ha flux
comes from gas ionized by massive O-stars and early-type B-stars, with
shorter lifetimes than those providing UV flux. As such, the
\Hans-derived SFR probes shorter time scales.

\citet{lee09} argue that the change of this relationship at
low SFRs may be due to stochastic variations in the SFH of low mass
galaxies, where the UV 
SFR averages over longer time scales and will typically be larger than
the \Ha SFR in the case of a bursty SFH. As shown in
Figure~\ref{hauv}, the BCDs (from our sample and from within LVL)
nearly always have larger \Ha SFRs than UV SFRs. This 
suggests that the BCDs may be in a 
phase of their SFH where intense star formation has only recently
begun, 
and is better 
 traced by the \Ha SFR 
than by
  UV. 
 The best-fit SFRs from the 
  SED fits are driven by the UV emission, and as such are not
  sensitive to the recent star formation 
   which produces \Ha emission. If 
the BCDs have had a recent enhancement in SF activity, the UV SFR
measurement could be diluted by the lower SF which may have preceded
the current event.

\subsection{Outliers and exceptional galaxies}
\label{individuals}

The comparisons from Section~\ref{external} between empirical
estimates of stellar
mass (via $3.6\mu$m and K$_s$ luminosity) and SFR (via 
\Ha and UV luminosity) with those from the SED fits were primarily 
intended as consistency checks of the SED fitting process. However,
they can also be used to test the reliability of different stellar
mass and SFR estimators for particular objects. For most galaxies in
our sample,
the SED fitting gives SFR and stellar mass values consistent with the
empirical methods.
%
%
However, the galaxies with inconsistent values
require further investigation. There can be many reasons for individual
galaxies to deviate from the normal relationships: they may be
intrinsically unusual objects in a brief evolutionary stage or an
uncommon region of parameter space, or may be examples of objects
which have SEDs that are not well fit by our methods. Here we look at
individual objects from the previous figures as examples of the types
of outliers in this work.




As shown in 
Section~\ref{external}, the SED-derived stellar masses are generally
in good agreement with the empirical estimates from $3.6\mu$m and
K$_s$ luminosity measurements. We have identified only very slight offsets
or non-unity slopes in these relationships. On the $3.6\mu$m plot in
the left panel of Figure~\ref{M}, there are no significant outliers
from the distribution of points. On the right panel of Figure~\ref{M}
which shows the stellar mass from K$_s$ photometry, the object
furthest above the relation is a BCD (UM~462) and the object furthest
below the relation is an LVL galaxy (UGC~01056). UM~462 has a
SED-derived stellar mass $\sim$$50\%$ larger than its K$_s$ mass, and
UGC~01056 has an SED-mass $\sim$$50\%$ smaller than its K$_s$
mass. These deviations are still within the Bayesian estimates of
uncertainty on the SED fits, and represent the most deviant mass
estimates. 

Throughout this analysis it has become clear that some members of our
BCD sample are less extreme than others. In particular, Mk~324,
Mk~328, and Mk~900 (labeled as ``BCD/E'' on the figures) typically
have properties more similar to the average LVL late-type
galaxies. Similarly, \citet{janowiecki14} 
found that these three were the 
reddest (B$-$V$\sim$$0.8$ in their outskirts), most luminous, and most
metal-rich objects in this
BCD sample. The SED fits have shown that these three also have the
largest stellar masses (M$_{\star}$$>$$4 \times 10^8 M_\odot$)
and the lowest specific star formation rates (SSFR$ < 3 \times
10^{-10}$ /yr) of the BCD sample. Given 
the dissimilarities between the three objects and the rest of the
BCD sample, we advise caution when drawing conclusions about BCDs
based on the inclusion of these three objects. They
have been included in our sample to cover the broad range of objects
commonly referred to 
as BCDs, but may be more intermediate between
extreme BCDs and more typical dwarf irregular galaxies.


\subsection{Future evolution of BCDs}

With their short gas depletion time scales and rapid star formation
rates, BCDs are in an unsustainable evolutionary state, and cannot
continue forming stars at their current rate. Many groups have
studied or discussed the implications of a population of faded or
post-burst BCDs \citep{sanchezalmeida08, sanchezalmeida09, amorin12,
  lelli12, koleva13, janowiecki14, meyer14}. Here it is important to
separate possible evolutionary pathways based on whether external or
internal effects are being considered.

\citet{meyer14} consider the long-term evolution of BCDs in the
Virgo Cluster, and find that BCDs may fade into galaxies similar to
the extant population of early type dwarfs in the Virgo Cluster. They
further suggest that in the cluster environment, a compact early type
dwarf may have its star formation re-ignited if it acquires fresh
gas. In lower density environments, \citet{sanchezalmeida08} searched
for BCDs in quiescence (QBCDs) by selecting galaxies with compact
structural parameters and other BCD-like properties, but without the
current intense star formation event. Within their search criteria,
they found a very large population of QBCDs, and suggest that perhaps
1/3 of all local dwarf galaxies may be capable of hosting a BCD-like
burst. Whether such intense starbursts are this common is still an
open question.

In this work, some of the SED-derived parameters for the BCDs and LVL
galaxies will evolve with time, while others will not. For example,
the initial burst fraction ($b_\textrm{frac}$) of the current star
star forming population will not decrease with time, while the young
population mass fraction will continue to decline. Similarly, the
total stellar mass will not change quickly, while the color will shift
from blue to red as the hottest youngest stars perish. We can generate
model predictions for the future evolution of BCDs, but this requires
that we assume there are no external effects driving this
evolution. 

In the simple case of internal evolution only, the stellar populations
of the BCDs would 
age normally, and follow the characteristic isochrones for
older ages. The resulting objects may resemble the so-called
``postburst'' galaxies \citep[or E+A, k+a, a+k,][]{dresslergunn83,
  dressler99}. These galaxies are characterized by strong Balmer line
absorption (from the most massive stars still alive, the A stars), but
no \Ha emission (as the O and B stars have perished). Other groups
have used the $4000$\AA \, break \citep{kauffmann03} or star formation
histories from resolved stellar populations
\citep{mcquinn10a,mcquinn10b} to identify galaxies which have
experienced strong star formation events. Further followup on
individual post-BCD-like galaxies is necessary to explore this
connection further.


\section{Summary}\label{sed_summary}

In this work we have used panchromatic observations (from UV to FIR)
of a large sample of 258 LVL galaxies and 18 BCDs to generate and fit
SEDs in order to determine key physical parameters about these
galaxies. These fits have allowed us to explore the role
that BCDs play in the larger context of dwarf galaxy evolution. In
particular, our main results are as follows:

\renewcommand{\labelitemi}{$-$}

\begin{itemize}

\item Our SED fits are able to reliably and robustly estimate key
  physical parameters of these galaxies, including M$_\star$
  and various SFRs.

\item SED model grids must be chosen to be broad and well-sampled
  enough to cover appropriate parameter space or else the best-fit
  values may have systematic offsets or errors.

\item 
Two-burst stellar population models fit the observed SEDs
  significantly better than one-burst models, suggesting that most or
  all dwarf galaxies have an old stellar population.

\item While powerful, the inability of SED fits to return a detailed
  SFH makes it difficult to answer questions about the evolutionary
  histories of individual BCDs.

\item 
When using \Ha SFRs in place of SED SFRs, the BCDs appear even more
extreme than typical galaxies, suggesting that the SF activity in BCDs
is uniquely intense or recent, and that SED fits are not
sensitive to the recent star formation traced by \Ha emission. 

\item We have identified unusual and potentially extreme objects for
  further study, which may represent brief evolutionary stages in
  dwarf galaxy evolution.

\end{itemize}

Further work is needed to more completely explore the evolutionary
relationships between
LVL galaxies and 
BCDs, and to connect the SED-fit results with the compactness
estimates of \citet{janowiecki14}. Once available, the optical
surface photometry of the LVL galaxies will provide an invaluable
resource for a more complete 
comparison between the BCDs and LVL galaxies. We will be able to
determine how extreme the BCDs are both in terms of the structural
parameters of their underlying hosts and in terms of their stellar
populations. 
BCDs appear to represent an extreme class of dwarf galaxy with
physical properties that place them at the edges of the broad
continuum of dwarf galaxy properties. 


\section{Acknowledgments}

SJ is grateful to D.~Cook for providing the LVL SEDs and
metallicities in a convenient format, and to S.~Salim and
H.~Evans for
useful discussions and advice.  We thank the anonymous referee for
helping to improve, clarify, and expand the context of this
work. All SED analysis was 
run on a laptop purchased with funds from a McCormick Science Grant
from the Indiana University College of Arts and Sciences. Some results
from this work were also presented in \citet{janowiecki15}. Support is
acknowledged from GALEX Cycle 3 program GI3-089.
SJ acknowledges support from the Australian Research Council's
Discovery Project funding scheme (DP150101734).

This research has made use of NASA's Astrophysics Data
System Bibliographic Services. This research has also made use of the
NASA/IPAC Extragalactic Database (NED), which is operated by the Jet
Propulsion Laboratory, California Institute of Technology, under
contract with the National Aeronautics and Space Administration.

Funding for SDSS-III has been provided by the Alfred P. Sloan
Foundation, the Participating Institutions, the National Science
Foundation, and the U.S. Department of Energy Office of Science. The
SDSS-III web site is http://www.sdss3.org.

SDSS-III is managed by the Astrophysical Research Consortium for the
Participating Institutions of the SDSS-III Collaboration including the
University of Arizona, the Brazilian Participation Group, Brookhaven
National Laboratory, University of Cambridge, Carnegie Mellon
University, University of Florida, the French Participation Group, the
German Participation Group, Harvard University, the Instituto de
Astrofisica de Canarias, the Michigan State/Notre Dame/JINA
Participation Group, Johns Hopkins University, Lawrence Berkeley
National Laboratory, Max Planck Institute for Astrophysics, Max Planck
Institute for Extraterrestrial Physics, New Mexico State University,
New York University, Ohio State University, Pennsylvania State
University, University of Portsmouth, Princeton University, the
Spanish Participation Group, University of Tokyo, University of Utah,
Vanderbilt University, University of Virginia, University of
Washington, and Yale University.

\appendix

\section{Scatter in the M-SFR relation}

We briefly consider systematic effects and biases across the M$_\star$
and SFR indicators used in this work. As a diagnostic, we start from
the assumption that galaxies on the star-forming main sequence
\citep[e.g.,][]{brinchmann04, salim07} populate a narrow range of star
formation rates at a given stellar mass. To assess the quality of of
our various M$_\star$ and SFR indicators, we can compare the 
dispersion about that relation using each combination of the values of
M$_\star$ and SFR produced in this work.

In this work, our M$_\star$ estimates include those using
simple mass-to-light ratios from monochromatic photometry (in both
$3.6\mu$m and K$_\textrm{s}$), and 
from our SED fitting analysis (e.g., Figure~\ref{M}). Our SFR
estimates include those from 
narrow-band \Ha photometry and from our SED
fitting (e.g., Figure~\ref{sfr}).

In order to consistently quantify the scatter in the M$_\star$-SFR
relation across different estimators, we first select LVL galaxies
which are not identified as early-types or BCDs. In this type of
comparison, galaxies with extreme (high or low) star formation
properties will add 
scatter to the main sequence M$_\star$-SFR relation, and should not be
included. Additionally, \citep{salim16} found that stellar mass 
determinations based on UV+optical+MIR SED fits are systematically
affected at different sSFRs by $\sim0.1$~dex (see their Figure~9). 

We also require that galaxies in this comparison have a complete set
of observations required to generate these estimates (i.e., fluxes
measured at $3.6\mu$m, K$_\textrm{s}$, \Hans, and a best-fitting SED
with reduced $\chi^2<5$). This yields a common sample of 97 LVL
galaxies which we use to compare M$_\star$ and SFR estimates.

\begin{figure*}
\centering
\includegraphics[width=\textwidth]{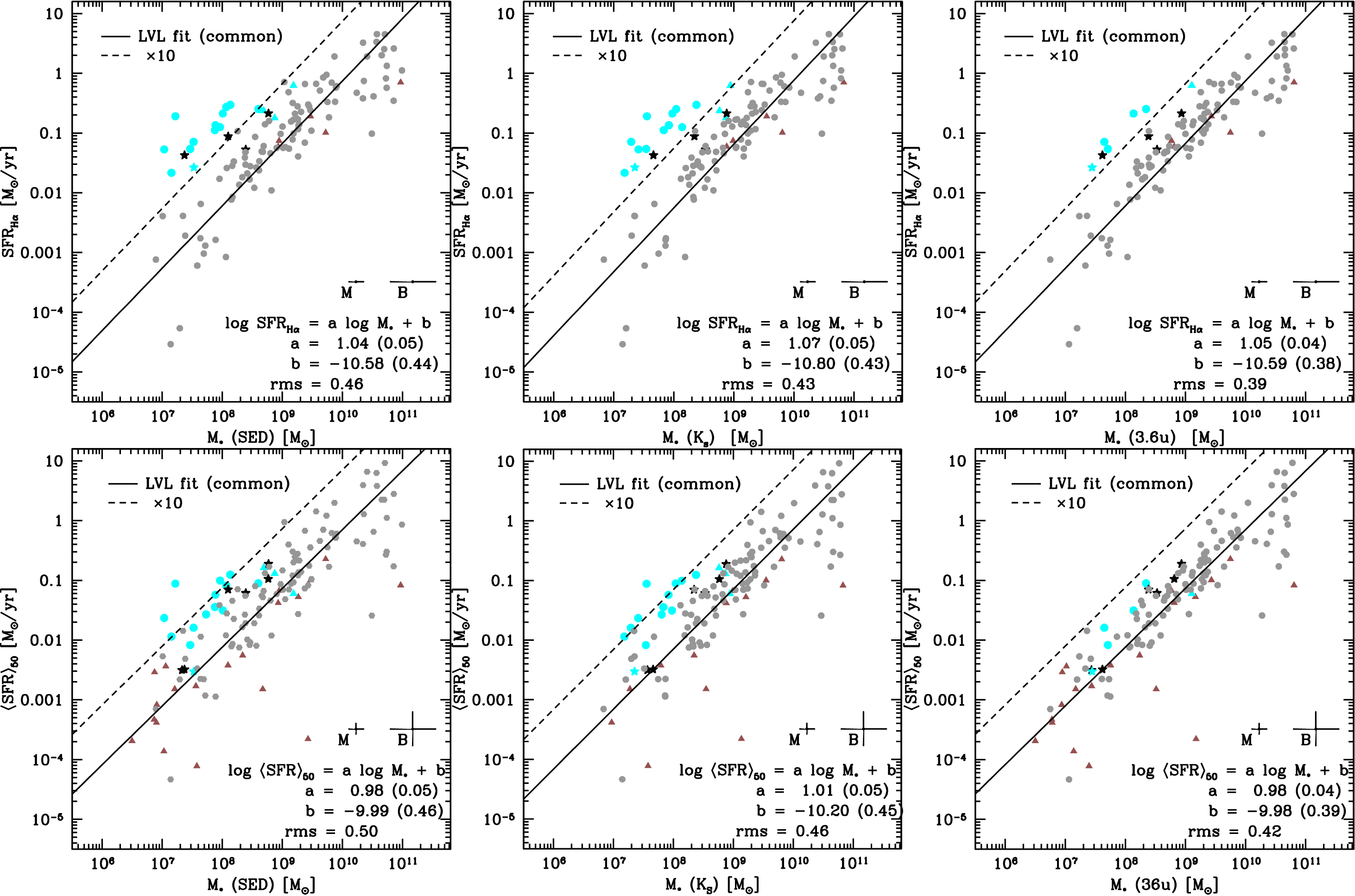}
\caption[rms test]{
Each panel shows SFR vs M$_{\star}$ for a common sample of 97
late-type (non-BCD) galaxies from LVL. This common sample is plotted
with grey dots and its best-fit parameters are given for linear least
squares fitting. Also shown (but not included in the fits) are members
of our BCD 
sample (cyan dots), our BCD/Es (cyan triangles), BCDs from the LVL
sample (black stars), and early-type galaxies from the LVL sample (red
triangles). Our results for Mk~475 are shown as a cyan star. The RMS
scatter about the main sequence is largest for the
SED-derived M$_{\star}$(SED) vs $\langle$SFR$\rangle_{50}$ relation,
and lowest for the M$_{\star}$(3.6$\mu$m) vs SFR$_{\textrm{H}\alpha}$
relation.
\label{rms}}
\end{figure*}

Figure~\ref{rms} shows the M$_\star$-SFR relations for each
combination of indicators in this common
sample. The top row 
compares the \Ha SFRs with our three estimates of mass (SED-fit,
K$_\textrm{s}$, and $3.6\mu$m), and the bottom row shows the
50-Myr-averaged SFR from our SED fitting. The results of linear
least squares fits to the SFR-M$_\star$ relations in each panel are
shown, and the RMS scatters are
given. Within this sample, the scatter about the fit increases from
M$_\star$($3.6\mu$m) to M$_\star$(K$_\textrm{s}$) to M$_\star$(SED);
and also increases from SFR(\Hans) to SFR(SED). The relation with the
lowest scatter (0.39~dex) is SFR(\Hans) vs M$_\star$($3.6\mu$m), and the
SED-fit quantities have the largest scatter (0.50~dex)

Given that the scatter in the SFR-M$_\star$ relation
increases when using SED fits, one might question whether the
SED-fitting is improving the determination of SFR and M$_\star$ in
this analysis. However, the goal of this project is not to simply
re-measure SFR and M$_\star$ in normal main sequence star-forming
galaxies using standard monochromatic indicators \citep[for works
  which focus on statistical samples of star-forming 
galaxies, see][and references therein]{brinchmann04, salim07,salim16}.
Rather, we are adopting a consistent SED-fitting methodology in order
to determine SFR and M$_\star$ even when considering more extreme
galaxies which have 
not previously been included in calibrating M$_\star$ and SFR
indicators. This approach allows us to quantify the differences
between BCDs (with extreme star formation) and the population of main
sequence star-forming galaxies, using consistent estimates of SFR and
M$_\star$.

Strong and/or recent star formation can bias monochromatic stellar
mass estimates of galaxies. For example, if a dwarf galaxy
undergoes a strong starburst, some 
of its K$_\textrm{s}$ and $3.6\mu$m luminosity will come from the
young stellar populations and those monochromatic estimates of stellar
mass could be systematically larger than for an otherwise similar
passive (or normal star-forming) galaxy.

Also note the relative positions of the four LVL galaxies classified
as BCDs (black stars) on the top row of the Figure~\ref{rms}. In each
panel those four galaxies have the same SFR(\Hans), but their M$_\star$
values are different. When using $3.6\mu$m or K$_\textrm{s}$
luminosity to estimate M$_\star$, these 
four galaxies are closer to the best-fit main sequence line for normal
galaxies. However, if the full SED-fitting is used, their M$_\star$
estimates are reduced -- it appears that their strong star formation
events may be artificially enhancing their $3.6\mu$m and K$_\textrm{s}$
stellar masses.

By including UV, optical, and IR photometry, our SED-fits produce more
robust M$_\star$ and SFR estimates for our diverse sample of galaxies
which have a wide 
range of star formation histories. This systematic and consistent
approach allows us to compare BCDs with a large sample of ``normal''
galaxies, without a bias toward main-sequence star-forming systems.

\end{document}